\documentclass[twocolumn,showpacs,preprintnumbers,amsmath,amssymb,showkeys]{revtex4}
\usepackage{graphicx}
\usepackage{dcolumn}
\usepackage{bm}

\begin{document}
\preprint{APS/123-QED}

\title{Peculiar properties of the cluster-cluster interaction\\
induced by the Pauli exclusion principle}
\author{Gennady Filippov}
\email{gfilippov@gluk.org}
\author{Yuliya Lashko}
\email{lashko@univ.kiev.ua} \affiliation{Bogolyubov Institute for
Theoretical Physics\\ 14-b Metrolohichna Str., Kiev-143, Ukraine}
\date{\today}

\begin{abstract}
Role of the Pauli principle in the formation of both the discrete
spectrum and multi-channel states of the binary nuclear systems
composed of clusters is studied in the Algebraic Version of the
resonating-group method. Solutions of the Hill-Wheeler equations
in the discrete representation of a complete basis of the
Pauli-allowed states are discussed for $^4$He$+n$, $^3$H$+^3$H,
and $^4$He$+^4$He binary systems. An exact treatment of the
antisymmetrization effects are shown to result in either an
effective repulsion of the clusters, or their effective
attraction. It also yields a change in the intensity of the
centrifugal potential. Both factors significantly affect the
scattering phase behavior. Special attention is paid to the
multi-channel cluster structure $^6$He+$^6$He as well as to the
difficulties arising in the case when the two clustering
configurations, $^6$He+$^6$He and $^4$He+$^8$He, are taken into
account simultaneously. In the latter case the Pauli principle,
even in the absence of a potential energy of the cluster-cluster
interaction, leads to the inelastic processes and secures an
existence of both the bound state and resonance in the $^{12}$Be
compound nucleus.
\end{abstract}


\maketitle

\section{Introduction}

Studying an interaction of the light nuclei,
it is necessary to take into account the Pauli exclusion
principle which significantly influences a behavior of the nuclei
at small distances between them. Thus the so-called "forbidden"
states, which can not be realized in reality, appear due to the
Pauli principle. Furthermore, the eigenvalues of the Pauli-allowed
states, which are not equal to unity, change a contribution of
the latter states to the wave function of the cluster system. At
last, matrix elements of the Hamiltonian and other operators
between allowed states contain their eigenvalues; and in this
respect they differ from the matrix elements of the same
operators, but calculated leaving out of the account the Pauli
principle.

There are several different approaches to taking into account the
Pauli exclusion principle in the problem of collision between
compound nuclear systems. The first approach (see, for instance,
\cite{Baz,Golb,Baz1,Zhuk}) takes into consideration the fact that
the Pauli principle does not allow identical nucleons to be at the
same point. And that is why a simulation of action of the
antisymmetrizer is realized by introducing an additional repulsive
potential between clusters. The choice of such potential is not
uniquely determined. As usual, in the approach stated above the
main requirement to be satisfied by a model potential is that it
should make a wave function of the nucleon system vanish,
provided the coordinates of the identical nucleons coinciding. In
the papers \cite{Baz,Golb,Baz1}, where the problems of
$\alpha$-$\alpha$ scattering and scattering of $\alpha$-particle
by the $^{12}$C, $^{16}$O, $^{20}$Ne nuclei have been considered,
the Pauli principle has been simulated with a phenomenological
repulsive potential of infinite intensity. The parameters of this
potential have been chosen to reproduce the data on the
continuous spectrum of the corresponding compound systems -- the
energies and widths of their resonance states. The authors of
paper \cite{Zhuk} simulated the Pauli principle for the systems
composed of several clusters within the method of hyperspherical
functions. As in \cite{Baz,Golb,Baz1}, they used a
phenomenological repulsive potential. But in this approximation
there is no guarantee that the forbidden states are excluded
completely, and a contribution of the allowed states meets all
the requirements of the Pauli principle.

To take into account action of the Pauli principle and to
eliminate the forbidden states, Saito suggested the orthogonality
condition model \cite{Saito}. In this model the allowed states
are found from the requirement of their orthogonality to the
forbidden states. In articles \cite{Kuk,Kuk1,Kuk2}
phenomenological potentials of a special kind were used. Such
potentials contain the operators of projection onto the forbidden
states and suppress these states by passage to the limit of high
intensity. To construct these potential, an explicit form of the
wave functions forbidden by the Pauli principle should be used.
The latter functions can be easily found, provided that the
interaction of the simplest nuclei being studied. As regards the
allowed states, they are not considered for constructing the
pseudopotentials; and, therefore, an influence of the Pauli
principle on these states is not taken into account, although it
may be essential.

Enumerated approaches turned to be fruitful and provided an
important information about discrete and continuum states of the
light nuclei composed of clusters. However, a fundamental problem
of constructing a complete basis of the states satisfying the
Pauli exclusion principle in the generator-parameter space and of
analyzing the influence of these states on the dynamics of
cluster systems (especially, multichannel ones) has not been
resolved yet.

It is worth mentioning that in all the cases of approximate
treatment of antisymmetrization effects the Pauli principle was
simulated with a repulsive potential, and its role has been
restricted to the elimination of the forbidden states from the
wave function. Meanwhile, an antisymmetrization operator affects
also Pauli-allowed states generating the matrix of an effective
potential which has a diagonal form in the representation of the
latter basis. By definition, an allowed state is an
orthonormalized eigenfunction of a norm kernel, provided that its
eigenvalue is non-zero. The eigenvalues of the norm kernel are
always positive. In case the eigenfunctions of the norm kernel
are normalized to the number of states, their eigenvalues show
how this normalization changes effectively under the influence of
the antisymmetrization operator. Change of the normalization of
the allowed states produces a direct effect on the cluster
dynamics.

The requirements of the Pauli principle can be accurately
fulfilled within the resonating-group method (RGM) \cite{Vild}.
But traditional form of the norm kernel and Hamiltonian kernel
commonly used for deducing the RGM dynamical equations
complicates analysis of the effects induced by the influence of
the Pauli principle, and thus many peculiarities of the cluster
behavior in collision are left beyond the scope of the study.

Detailed analysis of the exchange effects related to the
antisymmetrization can be performed with the use of the algorithm
outlined in \cite{Fewbody,Fewbody2}. Following the procedure
described there, we first define a complete discrete basis of the
harmonic-oscillator states allowed by the Pauli principle
(Section~\ref{sec:2}). Then we derive the Hill-Wheeler equations
in the representation of the discrete basis (Section
~\ref{sec:3}). General properties of the solutions obeying the
latter equations are discussed in Section~\ref{sec:4}.

In a consistent microscopic approach the forbidden states are
excluded, because their eigenvalues equal zero. And that is why
they do not enter an expression for the norm kernel, expansion of
which supplies a complete information on the spectrum of the
allowed states. Therefore, an effective potential related to the
antisymmetrization affects only the allowed states; and, as will
be shown later, it may not be a repulsive potential. A repulsion
arises in the states whose eigenvalues are less than unity,
whereas an attraction appears in the states with the eigenvalues
exceeding unity.

An algorithm of accounting the Pauli principle in calculation of
the matrix elements of the Hamiltonian and derivation of the
effective potential induced by the antisymmetrization operator is
suggested for a binary cluster system on the ground of the
generator-coordinate method and the Hill-Wheeler equation.
Properties of such a potential are analyzed and the conditions
fulfilling of which, by virtue of exact treatment of the
antisymmetrization effects, leads either to an effective
repulsion of the approaching clusters, or to their effective
attraction are established (Section~\ref{sec:5}).

By giving some examples of the binary cluster systems with one
open channel, a physical interpretation of the phenomena directly
related to an antisymmetrization of the wave function is
suggested (Section~\ref{sec:6}).

From the very beginning we use the allowed states only and do not
resort to the explicit form of the forbidden states neither to
introduce a pseudopotential nor to appeal to the orthogonality
condition model. Assuming a certain cluster structure of the
system studied or considering several cluster configurations
simultaneously, a complete basis of the Pauli-allowed states
(classified with the use of SU(3) symmetry indices and, in case
of the SU(3) degeneracy, additional quantum numbers)  is
considered along with a complete set of their eigenvalues. The
latter ones indicate an existence of the leading SU(3) irreps
that dominate in the continuum states with the low
above-threshold energy, and the amplitudes of which are maximal
for the discrete states of the cluster system. Basis states of
the leading representations have the largest eigenvalues.

Antisymmetrization effects in a binary cluster system with several
open channels are studied on the example of the continuum states
of $^{12}$Be that are able to decay through $^6$He$+^6$He and
$^8$He$+^4$He channels  (Section~\ref{sec:7}).

\section{\label{sec:2}Complete basis of the allowed states}

As  starting theses we shall use those which are initial for the
derivation of the Hill--Wheeler integral equations. At first, a
generating function of the system under consideration should be
determined. A form of this function depends on the degrees of
freedom, dynamics of which are of interest for us. These degrees
of freedom correspond to a complete basis of the allowed states.
Resort to the Hill--Wheeler integral equations implies the
transformation from the coordinate (or momentum) representation
to the representation of generator parameters.

Let $\Phi({\bf R},{\bf r})$ be the generator function of the
Hill--Wheeler method, antisymmetric with respect to a permutation
of the nucleon coordinates. Here ${\bf r}$ is the set of nucleon
vectors, ${\bf R}$ is the set of generator parameters describing
dynamics of the degrees of freedom which are of interest. We
construct this function as the Slater determinant composed of
nucleon orbitals to ensure its proper permutation symmetry. Such
function can be a linear superposition of the Pauli-allowed
states only. But we should show how to define a set of quantum
numbers $n$ for the basis of orthonormal states $\{\phi_n({\bf
r})\}$ generated by the function $\Phi({\bf R},{\bf r}).$ Also it
is necessary to find an explicit form of the basis functions. The
simplest way to do this is to introduce instead of the allowed
basis function $\{\phi_n({\bf r})\}$ defined in the coordinate
space its map $\{\psi_n({\bf R})\}$ in the generator parameter
representation. The simplification is attained due to the fact
that the number of generator parameters ${\bf R}$ is
significantly less than the number of single-particle variables
${\bf r}$ of the functions $\{\phi_n({\bf r})\}$
\footnote{Indeed, exactly the problem of reduction of the number
of independent variables of the wave function without violation
of the Pauli principle was resolved on the basis of the
generator-coordinate method suggested by Griffin and Wheeler
\cite{Wheel}.}.

In order to construct functions $\psi_n({\bf R})$, let us
introduce an expression
\begin{equation}
\label{x1} I({\bf S},{\bf R})=\int \Phi({\bf S},{\bf r})\Phi({\bf
R},{\bf r})d\tau,
\end{equation}
which is usually called the norm kernel or, in other words, the
overlap integral. Integration in (\ref{x1}) is over all
single-particle vectors. The norm kernel is symmetric with
respect to permutations of the generator parameters ${\bf R}$ and
${\bf S}.$ That is why it can be treated as a kernel of the
integral equation
\begin{equation}
\label{x2} \Lambda\psi({\bf R})=\int I({\bf S},{\bf R})\psi({\bf
S}^*)d\mu_{\bf S}.
\end{equation}
Symmetry of the kernel ensures an existence of its non-trivial
eigenfunctions $\psi_n({\bf R})$ and eigenvalues $\Lambda_n.$
Here $n$ is a set of quantum numbers of the basis functions. All
that remains is to define the integration domain of the generator
parameters ${\bf R}$ and ${\bf S}$ as well as the measure
$d\mu_{\bf R}$.

Both problems can be solved, provided the Slater determinant
$\Phi({\bf R},{\bf r})$ is composed of the Bloch-Brink orbitals
which are known to be the generating functions for the
single-particle harmonic-oscillator basis. The determinant
$\Phi({\bf R},{\bf r})$ needs a special discussion. For the first
time it was used as a trial function (but in a slightly different
form) in the $\alpha$-cluster Brink model, that is, in the
variational calculation of the spectrum of $^8$Be excited states.
Our determinant differs from the Brink determinant in two respects
\cite{Fewbody}. Firstly, it is generating function for the
many-particle harmonic-oscillator basis of the Pauli allowed
states. Secondly, its vector generator parameters take complex
values, and we treat them as independent variables of the allowed
states defined in the Fock-Bargmann representation.

If we restrict ourselves with one complex vector \[{\bf
R}=\frac{\bm{\xi}+i\bm{\eta}}{\sqrt{2}},\] where $\bm{\xi}$ and
$\bm{\eta}$ are real vectors, then an expression for the Bargmann
measure takes the form
\begin{eqnarray*}
d\mu_{\bf R}=\exp\{-({\bf R}\cdot {\bf
R}^*)\}\frac{d\bm{\xi}d\bm{\eta}}{(2\pi)^3}.
\end{eqnarray*}
The eigenvalues and eigenfunctions of the kernel (\ref{x1}) are
uniquely defined. Moreover, kernel (\ref{x1}) is a sum of
orthogonal degenerate kernels. Each of them corresponds to
definite values of the number of oscillator quanta $\nu$ and
SU(3) symmetry indices $(\lambda,\mu)$ that makes these kernels
be orthogonal. Therefore, solving of the equation (\ref{x2}) is
reduced to a standard algebraic procedure for the integral
equation with degenerate kernel. In the most general case and
only because the generator functions are constructed as the Slater
determinants composed of the Bloch-Brink orbitals, the
eigenfunctions of the norm kernel are labeled by the total number
of the oscillator quanta $\nu$, SU(3) symmetry indices
$(\lambda,\mu),$ an additional quantum number
$\alpha_{(\lambda,\mu)}$ when several different $(\lambda,\mu)$
multiplets exist, the angular momentum $L$, its projection $M,$
and, if necessary, one more additional quantum number
$\alpha_{L}$. The latter is needed to label the states with the
same $L$ in a given $(\lambda,\mu)$ multiplet. Then the
Hilbert--Schmidt expansion of the kernel of the integral equation
(\ref{x2}) is
\begin{equation}
\label{x11} I({\bf S},{\bf R})=\sum_n \Lambda_n\psi_n({\bf
S})\psi_n({\bf R}),
\end{equation}
where each of the eigenvalues $\Lambda_n$ of the norm kernel
corresponds to the eigenfunction $\psi_n({\bf R}).$ Naturally, the
eigenfunctions of the kernel (\ref{x1}) are orthogonal with
Bargmann measure and normalized to the dimensionality of the
irreducible representation $(\lambda,\mu)$ (irrep) \cite{Harvey}
\begin{eqnarray*}
\int d\mu_{\bf R} \psi_{(\lambda,\mu)}({\bf
R})\psi_{(\lambda,\mu)}({\bf R}^*)=
\frac{(\lambda+1)(\mu+1)(\lambda+\mu+2)}{2}.
\end{eqnarray*}

The second-order Casimir operator of the SU(3) group commutes with
the operator of permutation of the nucleon position vectors.
Hence, SU(3) symmetry indices naturally appear as the quantum
numbers of the eigenfunctions $\psi_n({\bf R}).$ Only such
quantum numbers ensure a diagonal form of the Hilbert--Schmidt
expansion. Any other classification of the basis states disrupts
the diagonal form of the expansion (\ref{x11}). For instance,
keeping $\nu$ as a quantum number, instead of the SU(3) symmetry
indices $(\lambda,\mu)$ quantum numbers of the angular
momentum-coupled ("physical") basis can be introduced. The states
of the latter basis (referred to as "$l$-basis" in what follows)
are labeled by the number of quanta $\nu$, the angular momenta of
each of the clusters $l_1$ and $l_2$, and the angular momentum of
their relative motion $l$ (see, for example,
\cite{Fewbody,Fewbody2}). Then unitary transform should be
applied to the functions $\psi_n({\bf R})$. But it results in an
off-diagonal form of the expansion (\ref{x11}), because all the
eigenvalues $\Lambda_n$ are not equal to unity. Only in the
asymptotic limit of the large number of oscillator quanta $\nu$
the eigenvalues appear to be close to unity; and the unitary
transformation from the SU(3) basis to the $l$-basis leaves the
expansion (\ref{x11}) intact, thus allowing to solve continuum
state problems in either basis \cite{Fewbody}.

In the absence of an SU(3) degeneracy the eigenfunctions of the
kernel (\ref{x1}) can be straightforwardly constructed with the
use of algebraic methods. In case of the SU(3) degenera\-cy the
eigenfunctions, along with their additional quantum numbers
$\alpha_{(\lambda,\mu)}$, are found by solving an integral
equation with the degenerate kernel (see, for example,
\cite{Fewbody2}), which is reduced to a set of homogeneous
algebraic equations with the rank equal to the degree of the
SU(3) degeneracy.

The dependence of the eigenvalues $\Lambda_n$ on the quantum
numbers of the allowed states is known. They take only positive
values, and they are the same for all the states with definite
SU(3) symmetry indices $(\lambda,\mu)$. In the presence of the
SU(3) degeneracy the eigenvalues for different SU(3) multiplets
with the same symmetry indices $(\lambda,\mu)$ do differ. At a
given $\nu$ we shall have several identical SU(3) irreps, i.e.
SU(3) degeneration. Then an additional index is necessary to
distinguish branches of the states with the same SU(3) symmetry.
At that, the eigenvalues belonging to the same branch should
change smoothly with $\nu$.

The eigenvalues appear to be nonzero starting with some minimal
number of quanta $\nu_{\textrm{min}}.$ As long as only the binary
cluster configuration with $\nu=\nu_{\textrm{min}}$ is
considered, there exists only one allowed SU(3) multiplet
$(\lambda_0,\mu_0)$ which corresponds to the Elliott's scheme for
the translation-invariant oscillator shell model spectrum
generated by the leading representation \cite{Ell}. But, in
addition to the orthonormalized basis of this scheme, we found
also the eigenvalue $\Lambda_{(\lambda_0,\mu_0)}.$ In case the
basis of the leading representation only is employed, this
eigenvalue has not practical influence on the results of
calculations. It comes into play when we invoke all the other
multiplets with $\nu>\nu_{\textrm{min}}$ along with their
eigenvalues instead of restricting ourselves with the multiplet
$(\lambda_0,\mu_0)$.

First, we dwell on the question of classification of the SU(3)
multiplets in the case of binary clustering confining ourselves
to the SU(3) irreps with even symmetry indices $\lambda$ and
$\mu$ \footnote{In other words, we shall not go beyond the states
having symmetry $A$ as $D_2$-symmetry.}. If $\nu=\nu_0+2,$ then
the allowed states belong to several (in the simplest case, two)
irreps: $(\lambda_0+2,\mu_0)$ and $(\lambda_1,\mu_1).$ Along with
these irreps, the eigenvalues $\Lambda_{(\lambda_0+2,\mu_0)}$ and
$\Lambda_{(\lambda_1,\mu_1)}$ also appear. With $\nu$ increasing,
the number of the Pauli-allowed SU(3) irreps grows to maximum
permissible value, which is the same for all $\nu$ starting with
$\nu_1.$

Irreps with the different number of quanta can be distributed
among several branches, with all the states of the same branch
having the same symmetry index $\mu$ and differing only in value
of the first index $\lambda$. That is, the irreps
$(\lambda_0+\nu-\nu_{\textrm{min}},\mu_0)$ are assumed to belong
to the first branch; while
$(\lambda_1+\nu-\nu_{\textrm{min}}-2,\mu_1)$, to the second one,
etc. Hierarchy among these irreps is established by the magnitude
of the eigenvalues $\Lambda_{\nu,(\lambda,\mu)}.$ The irreps with
the maximal values of $\Lambda_{\nu,(\lambda,\mu)}$ are leading
ones. In particular, the irreps
$(\lambda_0+\nu-\nu_{\textrm{min}},\mu_0)$ belong to the leading
irreps.

For many of the cluster systems (for instance, $^8$He+$^4$He and
$^6$He+$^6$He), although not for all, the least symmetric SU(3)
irreps correspond to the branch $(\lambda_0+\nu-\nu_0,\mu_0)$. As
for the most symmetric ones, they appear at $\nu=\nu_1.$

In a three-cluster (or multi-cluster) system the number of the
Pauli-allowed SU(3) irreps infinitely increases with $\nu.$

\section{\label{sec:3}Deduction of the Hill-Wheeler equations}

After the generating invariants $\Phi({\bf S},{\bf r})$ and
$\Phi({\bf R},{\bf r})$ having constructed, we can express the
Hamiltonian kernel of the cluster system under consideration in
terms of the basis functions as follows
\begin{eqnarray}
H({\bf S},{\bf R})&=&\int \Phi({\bf S},{\bf r})\hat{H}\Phi({\bf
R},{\bf r})d\tau \nonumber\\&=&\sum_n\sum_{\tilde{n}}\psi_n({\bf
S})<n|\hat{H}|\tilde{n}> \psi_{\tilde{n}}({\bf R}). \label{u1}
\end{eqnarray}
Let us express the wave function $\Phi({\bf r})$ of the
generator-coordinate method to be defined  as the Hill-Wheeler
integral (see, for example, \cite{Hor})
\begin{eqnarray*}
\Phi({\bf r})=\int C({\bf R}^*)\Phi({\bf R},{\bf r})d\mu_{\bf R},
\end{eqnarray*}
containing a new unknown function $C({\bf R}^*).$ The equation for
the latter follows from the variational principle for the
functional
\begin{equation}
\label{u10} \int\int C({\bf S}^*)\left[H({\bf S},{\bf R})-EI({\bf
S},{\bf R})\right] C({\bf R}^*)d\mu_{\bf S}d\mu_{\bf R}=0,
\end{equation}
where the energy $E$ makes sense of the Lagrange multiplier.

But here we shall deviate from a traditional deduction procedure
for the Hill-Wheeler equation. Instead, in order to reduce the
functional to an algebraic expression containing the expansion
coefficients $C^*_{n}$ ($C_{\tilde{n}}$) of an unknown functions
$C({\bf R}^*)$ $\left[C({\bf S}^*)\right]$ in the basis of the
Pauli-allowed states, we shall make use of the expansions
(\ref{x11}) and (\ref{u1}) for the norm kernel and the
Hamiltonian kernel, correspondingly, as well as the
orthonormality condition for the Pauli-allowed eigenfunctions.

Let
\begin{eqnarray*}
C({\bf R}^*)=\sum_n C^*_{n}\psi_{n}({\bf R}^*),\qquad C({\bf
S}^*)= \sum_{\tilde{n}} C_{\tilde{n}}\psi_{\tilde{n}}({\bf S}^*).
\end{eqnarray*}
Then
\begin{eqnarray}
\label{u11} \sum_n\sum_{\tilde{n}}
C^*_{n}\left(<n|\hat{H}|\tilde{n}>-E
\Lambda_n\delta_{n,\tilde{n}}\right)C_{\tilde{n}}=0.
\end{eqnarray}
Now we have two possibilities. Variation of the functional
(\ref{u11}) brings us to one of them that implies solving a set of
the algebraic equations
\begin{eqnarray}
\label{u12} \sum_{\tilde{n}}<n|\hat{H}|\tilde{n}>C_{\tilde{n}}-E
\Lambda_nC_{n}=0.
\end{eqnarray}
Certainly, $n$ takes all the values permissible for the
Pauli-allowed basis functions. Another possibility consists in a
diagonalization of the two quadratic forms containing in the
left-hand side of the equation (\ref{u11}).

\section{\label{sec:4}Solution of the Hill-Wheeler equation}

First, let us discuss general properties of solutions of the set
of equations (\ref{u12}). For a binary cluster system the
components of eigenvectors of the discrete states with the energy
$E_\kappa=-\kappa^2/2<0$ decrease exponentially with the number of
radial quanta $\nu=2k$ obeying the law
\begin{eqnarray*}
C_{n}^\kappa=A_n^\kappa\frac{\sqrt{2}\exp\left(-\sqrt{2\left|E_\kappa\right|}
\sqrt{4k+2l+3}\right)}{\sqrt{r_0}\sqrt[4]{4k+2l+3}}.
\end{eqnarray*}
Here $\{n\}$ are the quantum numbers of the $l$-basis, $l$ is the
angular momentum of cluster relative motion, $r_0$ is the
oscillator length, $A_n^\kappa$ is the asymptotic normalization
coefficient \cite{Bloch}.

Asymptotic behavior (at large values of the number of quanta
$\nu$) of the eigenvectors for the continuum states
$\left\{C_{n}(E)\right\}$ with  the energy $E>0$ is expressed in
terms of the Hankel functions of the first and second kind, and
the scattering $S$-matrix elements; or the Bessel and the Neumann
functions, and the $K$-matrix elements \cite{Fewbody,Fewbody2}.

The eigenvectors of the different states are orthonormalized, but
the orthonormality conditions contain the eigenvalues $\Lambda_n$
as the weight coefficients,
\begin{subequations}
\label{ort}
\begin{eqnarray}
\label{subeq:1} \sum_n\Lambda_n
{C_n^{\kappa}}^*C_n^{\kappa'}&=&\delta_{\kappa,\kappa'},\\
 \sum_n\Lambda_n {C_n}^*(E)C_n(E')&=&\delta(E-E'),
\label{subeq:2}\\
\sum_n\Lambda_n {C_n^{\kappa}}^*C_n(E)&=&0. \label{subeq:3}
\end{eqnarray}
\end{subequations}
A non-standard form of the orthonormality conditions (\ref{ort})
comes out from the fact that the coefficients ${C_n^{\kappa}}$ and
$C_n(E)$ are the eigenvectors of (\ref{u12}), the so-called
generalized eigenvalue problem \cite{Korn}.

The eigenvectors having been determined, it is easy to proceed
either to the eigenfunctions $\Phi_\kappa({\bf r}),$ $\Phi_E({\bf
r})$ in the coordinate representation or to the functions
$\Psi_\kappa({\bf R}),$
\\$\Psi_E({\bf R})$ in the Fock-Bargmann space. Then for the discrete
states we have
\begin{equation}
\label{wf_kappa} \Psi_\kappa({\bf R})=\sum_n
\sqrt{\Lambda_n}C_n^{\kappa}\psi_n({\bf R}),
\end{equation}
while for the continuum states we obtain
\begin{equation}
\label{wf} \Psi_E({\bf R})=\sum_n
\sqrt{\Lambda_n}C_n^{E}\psi_n({\bf R}).
\end{equation}
Now let us address another version of the procedure of
constructing the solutions of the Hill-Wheeler equation, namely,
to the diagonalization of the two quadratic forms in (\ref{u11}).
In order to reduce them to a diagonal form simultaneously, it is
worthwhile redefining the coefficients $C^*_{n}$ ($C_{\tilde
{n}}$) setting
\begin{eqnarray*}
\bar{C}^*_{n}=\sqrt{\Lambda_n}C^*_{n},\qquad \bar{C}_{\tilde
{n}}=\sqrt{\Lambda_{\tilde{n}}}C_{\tilde {n}}.
\end{eqnarray*}

Then, instead of (\ref{u11}), the following equation for the
coefficients is obtained
\begin{equation}
\label{u13} \sum_n\sum_{\tilde{n}}\left\{\bar{C}^*_{n}
\frac{\langle
n|\hat{H}|\tilde{n}\rangle}{\sqrt{\Lambda_n\Lambda_{\tilde{n}}}}
\bar{C}_{\tilde{n}}-E\delta_{n,\tilde{n}}
\bar{C}^*_{n}\bar{C}_{\tilde{n}}\right\}=0.
\end{equation}

After that it remains to make the renormalized matrix of the
Hamiltonian
\begin{equation}
\label{u4} \left |\left|\frac{\langle
n|\hat{H}|\tilde{n}\rangle}{\sqrt{\Lambda_n\Lambda_{\tilde{n}}}}
\right|\right|
\end{equation}
be diagonalized by means of the unitary transformation. Then an
expression for the density matrix in the Fock-Bargmann
representation can be written as follows
\begin{widetext}
\begin{equation}
\rho({\bf S},{\bf R})= \sum_\kappa\left\{\sum_n\sqrt{\Lambda}_n
{C_n^{\kappa}}^*\psi_n({\bf S})
\sum_{\tilde{n}}\sqrt{\Lambda}_{\tilde{n}} C_{\tilde{n}}^{\kappa}
\psi_{\tilde{n}}({\bf R})\right\}+ \int dE
\left\{\sum_n\sqrt{\Lambda}_n C_n^*(E)\psi_n({\bf
S})\sum_{\tilde{n}}\sqrt{\Lambda}_{\tilde{n}} C_{\tilde{n}}(E)
\psi_{\tilde{n}}({\bf R})\right\}. \label{dens}
\end{equation}
\end{widetext}
The summation in Eq.~(\ref{dens}) is over the discrete states,
while the integration is over the continuum states. This density
matrix gives us the information about the behavior of the density
distribution function in the phase space. Note, that the number
of independent variables in the expression (\ref{dens}) is
significantly less than in the distribution function $\rho(\{{\bf
r}\})$ defined in the coordinate space. This is the main
advantage of the generator-coordinate method, in general; and of
the Fock-Bargmann representation, in particular \footnote{Thus,
the Fock-Bargmann representation provides us with the simplest
means of realization of the generator-coordinate method and opens
new prospects of analysis of the distribution functions for the
cluster systems.}.

Having  integrated Eq.~(\ref{dens}) over the phase space, we can
reduce the diagonalized density matrix of the system of
interacting clusters to the following sum
\begin{eqnarray*}
\int\rho({\bf R},{\bf R}^*)d\mu_{\bf R}=\sum_\kappa \sum_n
\left|C_n^\kappa\right|^2\Lambda_n,
\end{eqnarray*}
provided that all the states of the system belong to the discrete
part of the spectrum.

Here
\begin{eqnarray*}
\left(C_n^\kappa\right)^2\Lambda_n=\left|\int \psi_n({\bf
R}^*)\Psi_\kappa({\bf R})d\mu_{\bf R}\right|^2
\end{eqnarray*}
is the realization probability of the state $\psi_n({\bf R})$ in
the wave function of the cluster system $\Psi_\kappa({\bf R})$.

So, employing basis of the Pauli-allowed states, we deal with a
discrete representation when for each state of the system,
belonging either to the discrete or continuum spectrum, its
eigenvector should be constructed. The latter is a set of the
expansion coefficients $\{\sqrt{\Lambda_n}C_n\}$ of the wave
function of this state in the basis of the Pauli-allowed states.
Absolute value of a coefficient squared gives us a realization
probability for the corresponding Pauli-allowed basis state; and a
convergence holds both for the bound states and continuum.

\section{\label{sec:5}How does an antisymmetrization operator act}

The RGM wave function belonging to the continuous spectrum of the
system composed of several clusters has a remarkably simple form
at large inter-cluster distances. In this region the potential
energy of cluster-cluster interaction can be neglected and there
is no need of taking into account an antisymmetrization operator.
Then the wave function is given by the product of the intrinsic
cluster wave functions and the wave function of free motion of
their center of masses.

With the inter-cluster distance decreasing, the wave function of
the system changes and is no longer reducible to the simple
product. Nucleons of different clusters are not isolated from
each other anymore and the region of their localization increases
that results in changing the internal cluster energies and the
energy of cluster relative motion, even if the potential energy
of cluster-cluster interaction is not taken into
consideration. 

Among the factors coming into play at this stage, the most
important are those which are directly related to the influence of
the Pauli exclusion principle. As a result, a general picture of
the phenomenon becomes intricate; and to interpret it completely,
each factor determining the final result should be carefully
analyzed. Further we shall concentrate our attention on the
phenomena directly conditioned by an antisymmetrization of the
RGM wave function and which reveal characteristic features of the
latter function at small inter-cluster distances. With that end
in view, some considerable simplifications will be introduced for
the description of the potential energy of cluster-cluster
interaction. Summarizing, if we take into consideration only the
Pauli-allowed states, we should answer, at least, the two
questions: (1) At which inter-cluster distances an interaction
generated by the Pauli exclusion principle appears? (2) What are
the main features of such interaction?

To understand the results of action of the antisymmetrization
operator, first, let us remind the above discussed set of the
algebraic equations (\ref{u12}) where only the operator of the
kinetic energy of the relative motion of clusters (in the c.o.m.
frame) is retained:
\begin{equation}
\label{syst_kin}
\sum_{\tilde{n}}<n|\hat{T}|\tilde{n}>C_{\tilde{n}} -E\Lambda_n
C_n=0.
\end{equation}
Due to a particular simplicity of the kinetic energy operator its
generating matrix element can be written in the form
\begin{eqnarray*}
T({\bf R},{\bf S})=\hat{T}_{\bf R}I({\bf R},{\bf S})=\hat{T}_{\bf
R}\sum_n \Lambda_n\psi_n({\bf R})\psi_n({\bf S}),
\end{eqnarray*}
where
\begin{equation}
\label{kin_op} \hat{T}_{\bf R}=-\frac{\hbar^2}{4mr_0^2}\left({\bf
R}^2-2({\bf R \cdot \bf \nabla_{\bf R} })-3+{\bf \nabla}_{\bf
R}^2\right)
\end{equation}
is the Fock-Bargmann map of the kinetic energy operator (see, for
example, \cite{Fewbody}). $m$ stands for the nucleon mass from
now on. Eq.~(\ref{kin_op}) is immediately followed by the
conclusion that the kinetic energy matrix  in the
harmonic-oscillator representation is tridiagonal, whence
\begin{eqnarray*}
T({\bf R},{\bf
S})=&&\sum_{\nu}\left\{\Lambda_{\nu-2}T_{\nu-2,\nu}\psi_{\nu-2}({\bf
R})+ \Lambda_{\nu}T_{\nu,\nu}\psi_{\nu}({\bf R})\right.\\&&\left.+
\Lambda_{\nu}T_{\nu,\nu+2}\psi_{\nu+2}({\bf
R})\right\}\psi_{\nu}({\bf S}).
\end{eqnarray*}
Here $T_{\nu,\tilde{\nu}}$ are the matrix elements of the kinetic
energy operator $\hat{T}_{\bf R}$ between the functions
$\psi_{\nu}({\bf R})$ normalized to unity with the Bargmann
measure,
\begin{eqnarray*}
T_{\nu,\tilde{\nu}}=\int d\mu_{\bf R}\psi_{\tilde{\nu}}({\bf
R}^*)\hat{T}_{\bf R}\psi_{\nu}({\bf R}).
\end{eqnarray*}
For simplicity, the single-channel case is considered here. For
such a case the basis functions differ only in the number of
oscillator quanta $\nu$.

The nonlocal kinetic energy operator is obtained. This is an
important stage in a procedure of deduction of a set of equations
for the coefficients $C_{n}.$ A typical equation of the set
(\ref{syst_kin}) looks like
\begin{eqnarray}
\Lambda_{\nu-2}T_{\nu,\nu-2}C_{\nu-2}&+&
\Lambda_{\nu}(T_{\nu,\nu}-E)C_{\nu}\nonumber \\&&+
\Lambda_{\nu}T_{\nu,\nu+2}C_{\nu+2}=0. \label{w1}
\end{eqnarray}
A standard system of the discrete representation which leaves out
of account the requirements of the Pauli principle,
\begin{eqnarray}
\label{free} T_{\nu,\nu-2}C_{\nu-2}+(T_{\nu,\nu}-E)C_{\nu}+
T_{\nu,\nu+2}C_{\nu+2}=0,
\end{eqnarray}
differs from the set of equations (\ref{w1}) in two aspects.
Firstly, only the matrix elements between the allowed states enter
the latter. Therefore, there is no need to introduce any
additional potential in the initial Hamiltonian neither to remove
the forbidden states, nor to make the wave function vanish as the
identical nucleons approach each other. Such a term would give
zero contribution in the set of equations (\ref{syst_kin}),
because the equations of our discrete representation are
constructed with these requirements being properly accounted for.

Secondly, elimination of the forbidden states still does not
resolve on the problem in toto and the matrix of the set
(\ref{w1}) is an evidence for such a statement. It contains the
eigenvalues $\Lambda_\nu$ of the Pauli-allowed states, and for
this reason it is not identical with the matrix (\ref{free}) of
the kinetic-energy operator of the cluster free motion.

Equations of the set (\ref{free}) are known to take asymptotic
form at large values of $\nu$  and to be reduced to the
Schr\"{o}dinger differential equation of free motion with
definite angular momentum  $l$ (see \cite{Fil:osc} for details).
Equations of the set (\ref{w1}) are of the same form at large
$\nu$ when all the eigenvalues $\Lambda_\nu$ equal unity. But at
such values of $\nu$, that the eigenvalues differ from unity and
start to decrease or to increase (remaining positive, of course),
asymptotic form of the equations becomes complicated; and they are
transformed into equations of motion in the field of the
potential generated by the antisymmetrizer. Our purpose is to
find these equations in order to reveal the main features of the
cluster-cluster interaction in their collision.

Equations of the set (\ref{w1}) for collision of the clusters in
the state with angular momentum $l$ can be written in the form of
the finite-difference equations
\begin{eqnarray*}
&-&{1\over2}\left\{\left(1+{\Lambda_{\nu-2}\over\Lambda_{\nu}}\right)
\left(\nu+{3\over2}-{(2l+1)^2\over8\nu}\right)+
1-{\Lambda_{\nu-2}\over\Lambda_{\nu}}\right\}\\
&\times&{1\over4}\left(C_{\nu+2}-2C_{\nu}+C_{\nu-2}\right)\\
&-&{1\over2}\left\{1+{\Lambda_{\nu-2}\over\Lambda_{\nu}}+
\left(1-{\Lambda_{\nu-2}\over\Lambda_{\nu}}\right)\left(\nu+{3\over2}-{(2l+1)^2\over8\nu}\right)
\right\}\\
&\times&{1\over4}\left(C_{\nu+2}-C_{\nu-2}\right)\\
&+&\left\{\left(1+{\Lambda_{\nu-2}\over\Lambda_{\nu}}\right)
{(2l+1)^2\over32\nu} \right.\\&&\left.
+{1\over4}\left(1-{\Lambda_{\nu-2}\over\Lambda_{\nu}}\right)\left(\nu+{1\over2}\right)
\right\}C_\nu= {mr_0^2\over\hbar^2}\,EC_\nu.
\end{eqnarray*}
The latter is transformed into the Bessel differential equation
in the limit $\nu\gg1$ when the eigenvalues $\Lambda_\nu$ can be
set to unity:
\begin{eqnarray}
\label{bes1} \left({d^2\over dy^2}+{1\over y}{d\over
dy}-{(2l+1)^2\over4}{1\over y^2}+{mr_0^2\over\hbar^2}\,2E\right)
C(y)=0;\\ \nonumber y=\sqrt{2\nu+2l+3}.
\end{eqnarray}
The diagonal matrix
$||U_{\nu,\tilde{\nu}}^{\textrm{Pauli}}\delta_{\nu,\tilde{\nu}}||$,
\begin{eqnarray}
U_{\nu,\nu}^{\textrm{Pauli}}=&&{1\over2}\left(1+{\Lambda_{\nu-2}\over\Lambda_{\nu}}\right)
{(2l+1)^2\over16\nu}\nonumber\\
&&+{1\over4}\left(1-{\Lambda_{\nu-2}\over\Lambda_{\nu}}\right)
\left(\nu+{1\over2}\right) \label{U_pauli},
\end{eqnarray}
can be considered as the matrix of the operator of the effective
cluster-cluster interaction generated by the Pauli principle.
Physical meaning of the first term in (\ref{U_pauli}) is quite
simple. It is the centrifugal potential that is renormalized with
the factor
$${1\over2}\left(1+{\Lambda_{\nu-2}\over\Lambda_{\nu}}\right).$$
The latter, along with the eigenvalues, tends to unity as
$\nu\rightarrow\infty.$ If the eigenvalues approach unity from
below as $\nu$ increases, the renormalization factor does not
exceed unity; therefore, partial suppression of the centrifugal
potential is observed. In other case, if the eigenvalues approach
unity from above, some strengthening of the centrifugal potential
occurs.

The second term,
\begin{eqnarray*}
{1\over4}\left(1-{\Lambda_{\nu-2}\over\Lambda_{\nu}}\right)
\left(\nu+{1\over2}\right),
\end{eqnarray*}
represents a finite-range potential generated by the Pauli
principle (referred to as "effective potential related to the
antisymmetrization" in what follows). Its intensity decreases in
magnitude as the difference $\Lambda_{\nu}-\Lambda_{\nu-2}$
vanishes. If the latter remains negative as $\nu$ grows (the
eigenvalues monotonically approach unity from above), then the
effective potential turns to be attractive. If the difference of
the eigenvalues remains positive, provided that $\nu$ increases
(the eigenvalues monotonically approach unity from below), then
the effective potential is repulsive. Obviously, the range of such
interaction depends on the width of the interval where the
eigenvalues deviate from unity.

Indeed, as was shown in Section~\ref{sec:4}, the wave function of
the binary cluster system can be presented in the form of the
expansion (\ref{wf_kappa}) or (\ref{wf}). If $\Lambda_{\nu}<1,$
the eigenvalues suppress the terms with small values of $\nu$ in
the expansions (\ref{wf_kappa},\ref{wf}) that can be naturally
interpreted as the action of effective repulsion forces at small
inter-cluster distances. This leads to a decreasing the ground
state energy by absolute value and to a corresponding change in
scattering phases due to an appearance of the additional effective
repulsion. But if $\Lambda_{\nu}>1,$ then the terms with small
${\nu},$ on the contrary, turn to be more preferred that can be
considered as an effective attraction.

So, as follows from all stated above, the antisymmetrization
effects result not only in the elimination of the forbidden
states, but also in changing the kinetic energy of cluster
relative motion as clusters approach each other. Let us begin
with the remark on the cluster behavior in the state with
definite angular momentum $l$ in the simple case, that any
cluster-cluster interaction is absent and influence of the Pauli
exclusion principle on the cluster motion is not considered. Then
a centrifugal potential is the only factor that changes the
velocity of the cluster relative motion in their collision. It
decreases the kinetic energy of the relative motion of the
clusters until they stop at the turning point
$r_{\nu}=r_0\sqrt{2\nu+2l+3}$ and then begin to fly away.

Antisymmetrization renormalizes intensity of the centrifugal
potential and this factor should be the first that affects
cluster velocity as clusters approach each other. With the
inter-cluster distance decreasing, the effective potential
related to the antisymmetrization also comes into play.

If the eigenvalue $\Lambda_\nu$ is monotone increasing function
of $\nu$, then some suppression of the centrifugal potential
occurs. As a result, velocity of the cluster relative motion
decreases a lesser degree than in the field of the centrifugal
potential of the free motion. But then clusters come within the
range of the effective potential related to the
antisymmetrization. In the case considered this potential
corresponds to the repulsion; and therefore, it decreases the
kinetic energy of cluster relative motion. If the eigenvalue
$\Lambda_\nu$ is monotone decreasing function of $\nu$, then some
strengthening of the centrifugal potential on the distances, that
the Pauli principle comes into play, slows down cluster motion a
greater degree than in the case of free motion. But after that
the relative velocity of the clusters increases due to the action
of the effective attractive potential.

It should be noted also that in the low-energy region the elastic
scattering phase is formed by the effective antisymmetrization
potential. As for changing a centrifugal barrier caused by the
Pauli principle, it affects, mainly, an asymptotic behavior of
the phase shift.

To understand what is the range of influence of the
antisymmetrizer on the structure of the wave functions, it is
appropriate to consider two limit values of the number of quanta,
$\nu_{\textrm{min}}$ and $\nu_{\textrm{max}}.$ As long as
$\nu<\nu_{\textrm{min}},$ there are no the Pauli-allowed states
belonging to the branch under consideration; and all the
eigenvalues equal zero. If the number of quanta satisfy the
inequality $\nu_{\textrm{min}}\leq\nu\leq\nu_{\textrm{max}},$
then the eigenvalues become positive, but deviate from unity.
Finally, provided that the inequality $\nu>\nu_{\textrm{max}}$
holds, the eigenvalues can be considered approximately equal
unity. The limit number of quanta $\nu_{\textrm{max}}$ is defined
in rather a relative way that demonstrates a diffuseness of the
antisymmetrization operator range.

The intensity of the effective cluster-cluster interaction induced
by the Pauli principle is determined by the natural combination of
the parameters (with the dimension of energy) entering the problem
considered,
$$E_x=\frac{\hbar^2}{mr_0^2(2\nu_{\textrm{max}}+2l+3)}.$$
Here $r_0\sqrt{(2\nu_{\textrm{max}}+2l+3)}$ defines the
inter-cluster distance where the antisymmetrization effects come
into play; and the quantity $E_x$ determines the energy range
where for the branches with $\Lambda_\nu>1$ one can expect
occurrence of the resonance phenomena caused by the influence of
the antisymmetrization potential or, at least, a maximum of the
scattering phase. Obviously, the larger is the value
$\nu_{\textrm{max}}$, the larger is the range of the
antisymmetrization potential, and the lower are the energies which
correspond to possible resonant states.

It is appropriate to compare the range of the antisymmetrization
operator with that of the cluster-cluster interaction generated by
the nucleon-nucleon potential. At large values of the number of
quanta $\nu$ the matrix of the potential energy operator $U(r)$
of cluster-cluster interaction is known to be equivalent to the
diagonal matrix
$||\delta_{\nu,\tilde{\nu}}U(r_0\sqrt{2\nu+2l+3})||
$
(see \cite{osc})
that significantly simplifies the
above-mentioned comparison. For a central nucleon-nucleon
potential having a Gaussian form we will have
\begin{equation}
\label{range_nucl}
U(r_0\sqrt{2\nu+2l+3})=U_0\exp\left\{-\frac{2r_0^2(2\nu+2l+3)}
{b_0^2}\right\},
\end{equation}
where $b_0$ is the radius of the Gaussian potential, and $U_0$ is
its intensity.

At large values of the number of oscillator quanta $\nu$ the
effective cluster-cluster interaction induced by the
antisymmetrization vanishes as the eigenvalues $\Lambda_\nu$ tend
to unity. Thus, the range of such interaction can be estimated
with the help of the relation
\begin{eqnarray}
\Lambda_{\nu}-1\sim \beta(\nu)\,\alpha^{-\nu}
=\beta(\nu)\exp\left( -\nu\ln\alpha\right). \label{range_eff}
\end{eqnarray}
Parameter $\alpha>1$ is completely determined by the type of
clustering of the nuclear system considered; and that is why it is
the same for all the branches of the SU(3) irreps. As for the
parameter $\beta(\nu)$, its dependence on the number of quanta
obeys a power law and differs for different SU(3) irreps even
within one cluster configuration.

It follows from (\ref{range_eff}) that the effective interaction
induced by the antisymmetrization decreases exponentially with
the number of quanta, and so is the cluster-cluster interaction
generated by the nucleon-nucleon forces. Comparing the decrements
of the expressions (\ref{range_nucl}) and (\ref{range_eff}), it is
possible to establish which of the potentials has larger range.
It appears to be that
\begin{eqnarray*}
\frac{4r_0^2}{b_0^2}\gg \ln\alpha
\end{eqnarray*}
even for the simplest binary cluster systems composed of
$s$-clusters, such as $^4$He+n, $^4$He+$^4$He, etc. This means
that as the clusters approach each other, they first experience an
influence of the effective interaction caused by the Pauli
exclusion principle. And the potential of the cluster-cluster
interaction generated by the nucleon-nucleon forces comes into
play only at the smallest inter-cluster distances.

Further, on the examples of different binary systems composed of
$s$-clusters and $p$-clusters we shall demonstrate how the
eigenvalues (and, therefore, the parameters of the effective
cluster-cluster interaction induced by the antisymmetrizer)
depends on the number of nucleons of compound system and on the
type of clustering being considered.

\section{\label{sec:6}Examples of the effective potentials: single-channel cluster systems}

Our immediate purpose is to demonstrate the influence of the
effective potential related to the antisymmetrization on the
behavior of scattering phases and wave functions in the discrete
representation (expansion coefficients of the wave functions in
the complete basis of the Pauli-allowed states) addressing the
simplest examples of the cluster systems with one open channel.
We will also draw attention to the comparison of our results and
those which can be obtained by simulation of the Pauli principle
with some phenomenological potentials of the optical model.

Formally, the equations of the optical model with forbidden states
for the interaction of compound systems \cite{Kuk2} are the most
similar to our approach. In this model an antisymmetri\-zation is
performed by the aid of the procedure of elimination of the
forbidden states, for what a pseudopoten\-tial is introduced in
the equations.
In our equations the forbidden states are excluded from the very
beginning, because their eigenvalues are equal to zero, i.e.
$\Lambda_\nu=0$ for all $\nu<\nu_{\textrm{min}}$. The known
asymptotic estimate which establishes a relation between the wave
function $\Phi_l({\bf q})$ in the coordinate representation and
coefficients $C_{\nu}^{l},$
\begin{eqnarray*}
\Phi_l(|{\bf
q}|=r_0\sqrt{2\nu+3})={\sqrt{\Lambda_\nu}C_\nu^l\over{\sqrt{2r_0}(2\nu+3)^{1/4}}};\qquad
\nu\gg 1,
\end{eqnarray*}
is followed by the conclusion that the wave function $\Phi_l$ is
rather close to zero within the interval $$0<|{\bf
q}|<r_0\sqrt{2\nu_{\textrm{min}}+3}.$$ But such a wave function is
obtained in the calculations with a model repulsive potential in
the form of the core (see, for example, \cite{Baz,Baz1}). Our
conclusion is that the core should not be hard, because in such a
case the scattering phase would infinitely increase by absolute
magnitude with the energy. But the phase actually vanishes at
high energy, as it should be for the Born approximation.
Therefore, the model core should be soft; and its intensity and
radius should depend both on the number of nucleons in clusters
and on the angular momentum $l$ of cluster relative motion.

The number of the forbidden states with angular momentum
$l<\nu_{\textrm{min}}$ equals $[(\nu_{\textrm{min}}-l)/2].$
Hence, the elastic scattering phase $\delta_l(E)$ is appropriate
to be counted off from $[(\nu_{\textrm{min}}-l)/2]\pi,$ in order
to make it vanish at high energy. Such energy dependence of the
scattering phase can be reproduced by means of the soft core,
with its height being determined by the phase amplitude.
Naturally, the larger is the number $(\nu_{\textrm{min}}-l)/2$ of
the forbidden states, the larger should be the height of the core
and its radius. Therefore, the radius of the core depends not
only on $\nu_\textrm{min},$ but also on the angular momentum $l$;
and it can be estimated with the help of the relation
$R=r_0\sqrt{2\nu_{\textrm{min}}+3}$.

Note, that for the states with angular momentum $l\geq
\nu_{\textrm{min}}$ the forbidden states are absent, and hence
any core need not be introduced. Certainly, the larger is
$\nu_{\textrm{min}}$, the more accurate are the aforementioned
considerations.

But all stated above does not mean that a core reproduces in toto
the action of the Pauli principle. As was explained in the
previous section, taking into account the Pauli principle leads
to an appearance of the eigenvalues of the allowed states. They
deviate from unity at $\nu>\nu_{\textrm{min}}$ and affect the RGM
Hamiltonian in discrete basis representation. In particular, some
additional terms generated by the antisymmetrization operator
enter the Hamiltonian; and their role depends on changing the
eigenvalues of the Pauli-allowed states with the number of quanta.

The cluster systems considered in this section have a remarkable
property: any given number of quanta $\nu$ corresponds with only
one SU(3) multiplet $(\nu,0)$. That rather simplifies an explicit
form of the Pauli-allowed orthonormal basis functions in the
Fock-Bargmann space (they are constructed from even powers of one
complex vector) and provides only two sets of the eigenvalues: one
for the even states and the other, for the odd ones. Besides, the
matrix elements of the kinetic energy operator take a
particularly simple form in this case. Really, all the allowed
states belong to the branch of SU(3) irreps with
$(\lambda,\mu)=(\nu,0);$ and so the kinetic energy matrix, in
addition to the diagonal elements, contains only those which
couple SU(3) irreps $(\nu,0)$ and $(\nu+2,0).$

Now our purpose is to make the situation under consideration more
realistic and, at the same time, reveal all that directly relates
to the antisymmetrization. Towards this end,  we shall discuss the
results obtained in the approximation of zero-range nuclear
forces, which takes into account a potential energy operator only
in the most compact configuration for each cluster system; or
without this operator at all when trying to emphasize the role of
the antisymmetrization.

Behavior of the wave functions in discrete representation at the
relatively small number of quanta (where the influence of the
antisymmetrization is the most noticeable) is of special interest.
As to the phases of the elastic scattering, we concentrate our
attention on the region of energies up to 50 MeV (in the c.o.m.
frame). In the absence of the operator of potential energy of the
nucleon-nucleon interaction the two different modes in a behavior
of the scattering phases are observed in this region. Later we
shall see that at low energies the eigenvalues approaching unity
from above are correspondent with a positive scattering phase;
while those approaching unity from below, with a negative one.

\subsection{System n+$^4$He}

The scattering of a neutron and $\alpha$-particle is one of the
simplest examples considered in the frame of the Hill-Wheeler
method. We aim at giving a demonstration that even for this
system the scattering phase and the wave function can not be
reproduced in full by any optical model potential which simulates
an action of the Pauli principle.

The norm kernel for $\alpha$+n system generating a complete basis
of the Pauli-allowed states or, in other words, the kernel of an
integral equation looks like
\begin{equation}
\label{z1} I_{\textrm{n}+\alpha}({\bf R},{\bf S})= \exp({\bf
R}\cdot{\bf S})-\exp\left[-\frac{1}{4}({\bf R}\cdot{\bf
S})\right].
\end{equation}
The norm kernel does not contain the Pauli-forbidden states. They
are already excluded. Therefore, it can be compared with the
result of action of  the model repulsive potential which
eliminates the forbidden states. But we should also take account
of the eigenvalues of the allowed states. They are not equal to
unity and enter both the norm kernel and the matrix elements of
the Hamiltonian between the allowed states. Therefore, provided
that the Pauli principle is simulated by some phenomenological
potential, at this stage of solving the problem there should be
introduced some additional interaction to specify dynamics of the
system after elimination of the forbidden states. Later we will
attend to the question what should be the main features of such
an additional phenomenological potential.

Let us start with calculating the eigenvalues. Eq.~(\ref{z1}) is
immediately followed by the Hilbert--Schmidt expansion
\begin{equation}
\label{kern_alphan} I_{\textrm{n}+\alpha}({\bf R},{\bf S})=
\sum_{\nu=1}^\infty\left\{1-\left(-\frac{1}{4}\right)^\nu\right\}\frac{1}{\nu!}
({\bf R}\cdot{\bf S})^\nu.
\end{equation}
The expression
\begin{eqnarray*}
I^{(\nu,0)}({\bf R},{\bf S})=\frac{1}{\nu!}({\bf R}{\bf S})^\nu
\end{eqnarray*}
is the norm kernel for the irrep $(\nu,0)$, which is normalized
to the number of states. It can also be considered as a kernel of
the integral equation
\begin{eqnarray*}
\psi_{\nu,L,M}({\bf R})=\int I^{(\nu,0)}({\bf R},{\bf
S})\psi_{\nu,L,M}({\bf S}^*) d\mu_{\bf S}.
\end{eqnarray*}
At that, all the eigenvalues of the kernel $I^{(\nu,0)}$ are equal
to unity for any given quantum numbers $\nu,L,M$. That is why the
Hilbert--Schmidt expansion of the norm kernel (\ref{kern_alphan})
takes the form
\begin{eqnarray*}
I_{\textrm{n}+\alpha}({\bf R},{\bf S})=&&
\sum_{\nu=1}^\infty\left\{1-\left(-\frac{1}{4}\right)^\nu\right\}\\&&\times
\sum_{L,M}\psi_{\nu,L,M}({\bf R}) \psi_{\nu,L,M}({\bf S}).
\end{eqnarray*}
Therefore, the eigenvalues of the norm kernel for $\alpha$+n
system are equal to
\begin{eqnarray*}
\Lambda_{\nu}=1-\left(-\frac{1}{4}\right)^\nu.
\end{eqnarray*}
Obviously, in the limit $\nu\to\infty$ the eigenvalues tend to unity from below, if the number
of quanta is even ($\nu=2k$); and from above, if the number of
quanta is odd ($\nu=2k+1$)
\begin{eqnarray*}
\Lambda_{2k}=1-\left(\frac{1}{16}\right)^{k},\qquad
\Lambda_{2k+1}=1+\frac{1}{4}\left(\frac{1}{16}\right)^{k}.
\end{eqnarray*}
The minimal number of quanta $\nu_{\textrm{min}}$ which
corresponds to the lowest Pauli-allowed basis state is equal to
1. It means that the branch which belongs to the SU(3) irrep $(2k+1,0)$
appears first and its eigenvalues take the highest possible
values.

\begin{figure*}[e]
\includegraphics[width=8.6cm]{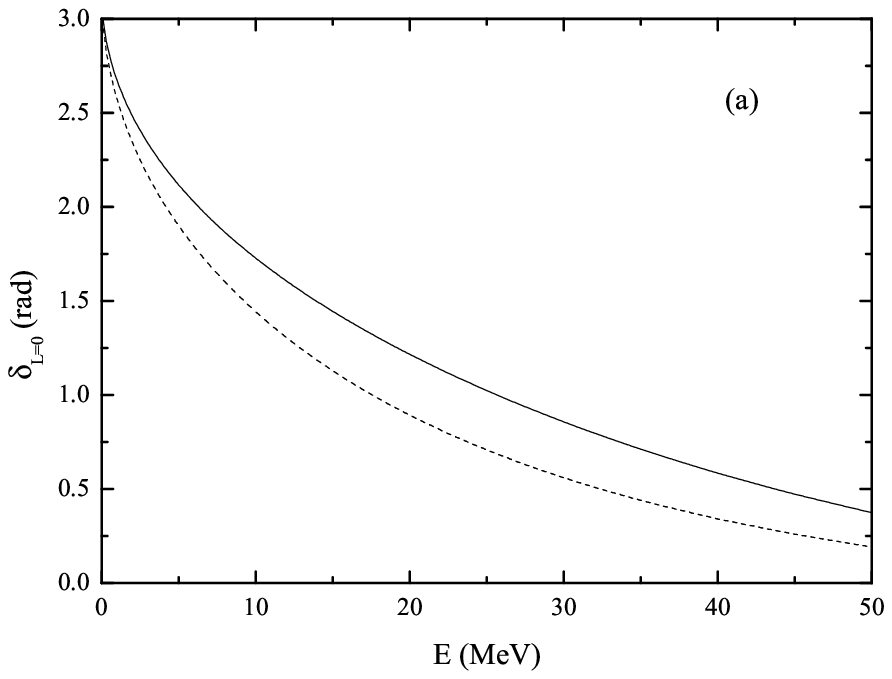}
\includegraphics[width=8.6cm]{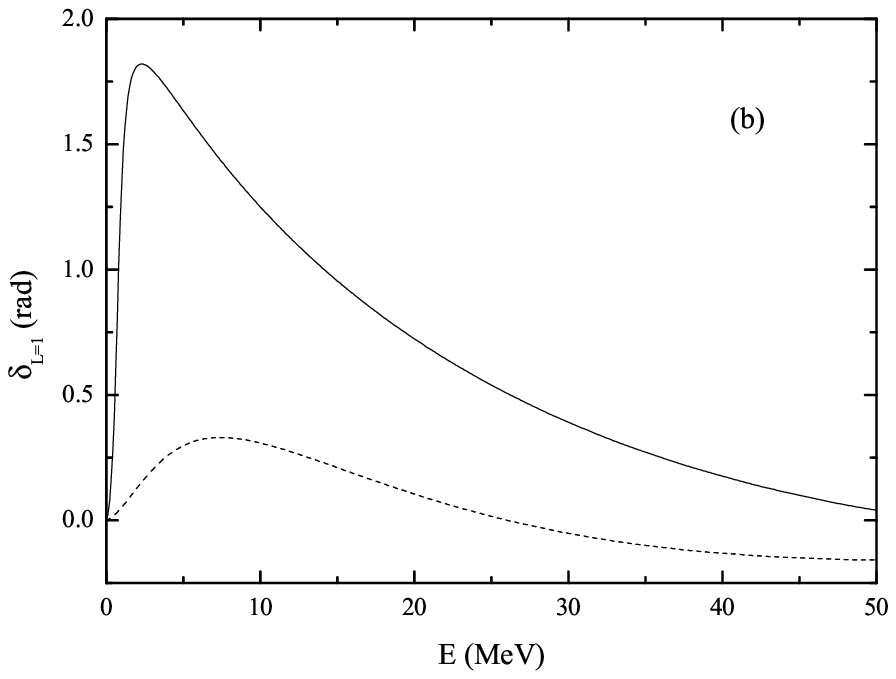}
\caption{\label{fig:1} Phases of the $\alpha$+n scattering for
the states with (a) $L=0$ and (b) $L=1$. Solid curve: phases
obtained in the approximation of zero-radius for nuclear force.
Dashed curve: phases obtained by granting the Pauli principle
only.}
\end{figure*}

To demonstrate a validity of the stated above conclusions about
the characteristic features of the effective interaction induced
by the Pauli principle on the concrete examples, it is appropriate
to consider the phases of the elastic scattering of a neutron by
$\alpha$-particle \footnote{These phases have already been
calculated earlier \cite{Tan}, but our purpose is to show how and
to what extent they are formed under the influence of the
antisymmetrization operator. We should also answer a question
whether it is possible to simulate the antisymmetrization effects
by the repulsive potential in the form of a soft or hard core.}
in the states with $L^{\pi}=0^+$ [Fig.~\ref{fig:1} (a)] and
$L^{\pi}=1^-$ [Fig.~\ref{fig:1} (b)]. Zero angular momentum
corresponds to the eigenvalues $\Lambda_{2k}<1$, while the
eigenvalues $\Lambda_{2k+1}$ which exceed unity correspond to the
momentum $L=1$.
The behavior of the scattering phases $\delta_{L=0}$ and
$\delta_{L=1}$ for the case when only an antisymmetrization is
taken into account and in the approximation of zero-radius for
nuclear force is presented in Fig.~\ref{fig:1}. In this
approximation the potential energy of the interaction of the
neutron and $\alpha$-particle is simulated with the single
parameter, the diagonal matrix element of the potential energy
operator ${\hat U}$ in the state with the minimal number of even
(for $L=0$) or odd (for $L=1$) quanta, i.e.
\begin{widetext}
\begin{eqnarray*}
\langle(\nu,0)|U|(\nu',0)\rangle =\left\{
\begin{array}{cc}U_0=- 5.52 \text{ MeV} & \text{ if }
\nu=\nu'=\nu_{\textrm{min}}=2,~L=0 \\\bar{U}_0=- 11.05 \text{
MeV} & \text{ if } \nu=\nu'=\nu_{\textrm{min}}=1,~L=1
\\0  & \text{ otherwise }
\end{array} \right.
\end{eqnarray*}
\end{widetext}
The parameter $\bar{U}_0$ was fitted to reproduce a position of
the maximum of the total cross-section of the elastic scattering
$E_r=0.92\pm0.04$ MeV \cite{Ajzenberg}. The half-width
$\Gamma=1.3$ MeV also agrees well with the experimental value of
$\Gamma=1.2$ MeV, although the maximum value of the total cross
section twice as large as its experimental value of 7.6
barns \cite{Ajzenberg}. As regards the parameter $U_0$, it was
chosen to provide a reasonable description of the experimentally
observed phase of the $\alpha$+n elastic scattering with $L=0$ at
low energies \cite{alphan}.

As long as the energy $E$ of the relative motion of clusters is
small, the phase of the elastic scattering with angular momentum
$L$ obeys the law
$$\delta_L\sim \pi n-a_L (2E)^{L+1/2},$$
i.e. as for a standard short-range potential. Here $n$ is the
number of bound states. If $L=0,$ then the factor $a_0$ is called
the scattering length. There are no bound states in the system
$\alpha+$n, but there exists one Pauli-forbidden state at $k=0$
with angular momentum $L=0$, which is known to have the same
influence on the behavior of the scattering phase $\delta_0$ as a
bound state \cite{Saito77}. Therefore, at zero energy the
scattering phase $\delta_0$ is naturally counted off from $\pi.$
For the odd number of quanta there are no forbidden states and we
set the scattering phase $\delta_1$ be equal to zero at zero
energy.

The positive sign of the scattering length conforms to the known
general consideration that Pauli principle can be simulated by a
repulsive potential. Of course, the attraction $\bar{U}_0$
appeared not to be strong; otherwise it would change the signs of
the scattering phase and scattering length.

As it can be seen from Fig.~\ref{fig:1} (b), $a_{L=1}<0$.
Therefore, an effective interaction caused by the Pauli principle
for the states with angular momentum $L=1$ is attractive. Note
that in the low-energy range up to 20 MeV the scattering phase is
positive both for the case with potential $\bar{U}_0$ and with the
antisymmetrization effects being taken into account only. We
comparing the scattering phases with angular momentum $L=1$
without a potential and with the potential $\bar{U}_0$, the
intensity of the attraction induced by the antisymmetrization
effects only is not high enough to assure an existence of the
experimentally observed $1^-$-resonance in the continuum of the
$^5$He nucleus. But a contribution from the antisymmetrization is
not neglible, because the range of the potential induced by the
Pauli principle exceeds that of the nuclear forces.

Stress that the energy dependence of the scattering phase
$\delta_1$ obtained with the account of antisymmetrization
effects can not be reproduced by simulation of action of the
Pauli principle with a soft or hard core. Such an approximation
would be
bad, because it could not explain a positive sign of the phase at
low energy.

Now let us resort to the expansion coefficients of the wave
functions of the continuous spectrum at different energies to
understand the behavior of these functions in different states.
This opens a prospect for a more detailed investigation of the
structure of the states for which only the energy dependence of
the scattering phases has been available so far.

\begin{figure*}[e]
\includegraphics[width=8.6cm]{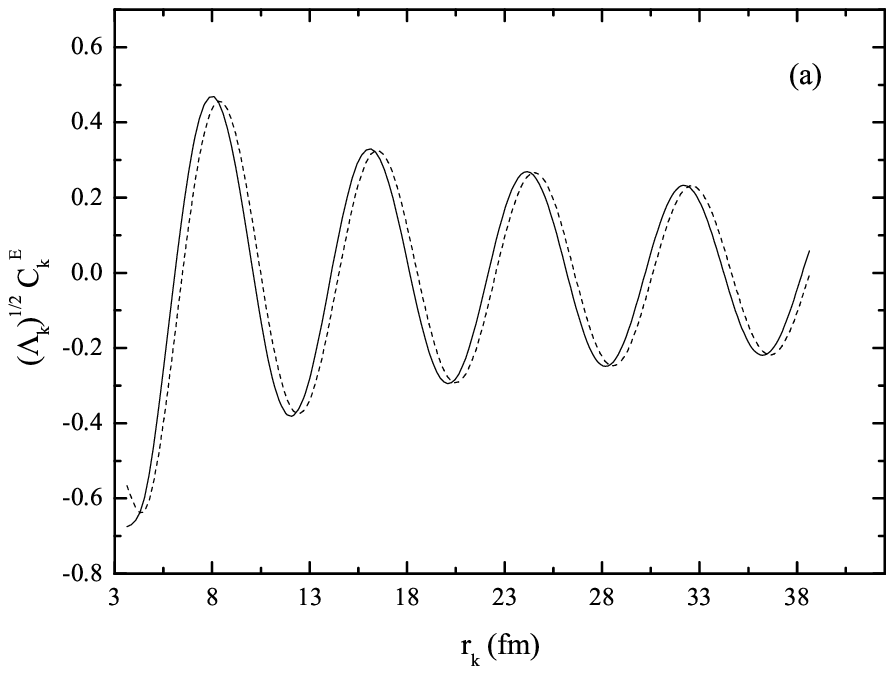}
\includegraphics[width=8.6cm]{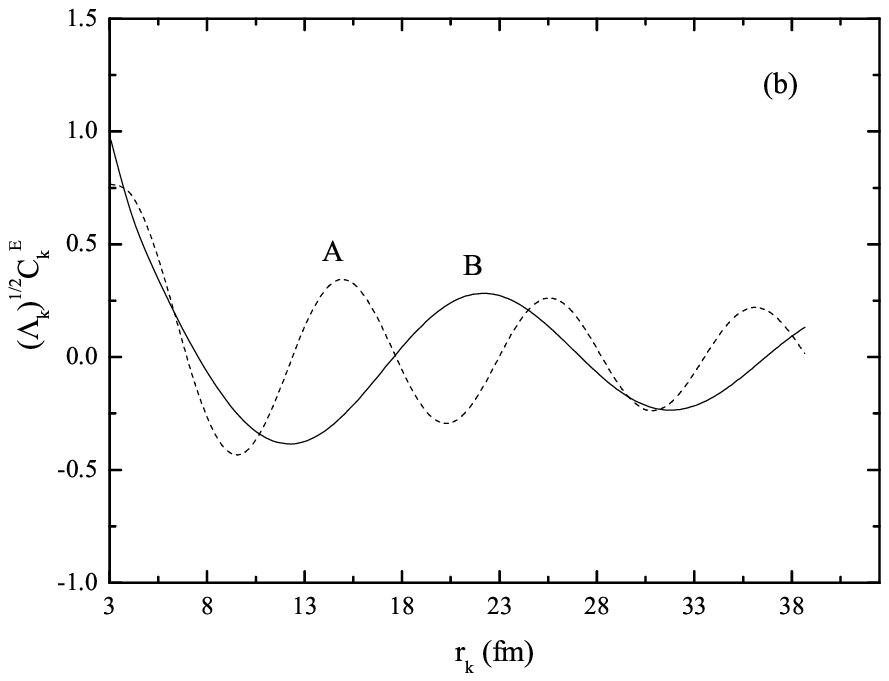}
\caption{\label{fig:2} Continuum states of $^5$He: coefficients of
the w.f. expansion in the SU(3) basis
 (a) $L=0$, $E=12.61$ MeV and (b) $L=1$, line A: $E=7.41$ MeV, line B: $E=2.31$ MeV.
 Solid curve: w.f.
obtained in the approximation of zero-radius for nuclear force.
Dashed curve: w.f. obtained by granting the Pauli principle only.}
\end{figure*}

First, let us consider the wave functions of the channel
$L^\pi=0^+.$ Wave functions at the energy $E=12.61$ MeV (at this
energy the scattering phase calculated in the approximation of
zero-range for nuclear force equals to $\pi/2$) are shown in
Fig.~\ref{fig:2} (a) for the variants with and without the
potential $U_0$. This value of the scattering phase is notable
for the absence of the Bessel function in a corresponding
expression for the wave function at large value of the number of
quanta $k$; only the Neumann function
$N_{1/2}(\sqrt{2E}r_0\sqrt{4k+3})$ remains there.
The wave functions for both versions almost do not differ except
a small shift of the wave to the origin in the case with the
potential. This shift is induced by action of the weak attractive
potential $U_0.$

The wave function of the $L^\pi=1^-$ state at the energy
corresponding to the maximum value of the scattering phase is also
of interest. As is well known, taking into account a spin-orbit
interaction leads to the splitting of the state $L^\pi=1^-$; and,
as a result, the two scattering phases $\delta_{3/2}$ and
$\delta_{1/2}$ appear, each of them showing a resonance behavior.
In our case it is possible to conclude about only one resonance
located between the mentioned above. The phase $\delta_1$
calculated in the version with the potential $\bar{U}_0$ reaches
its maximum value at $E=2.31$ MeV. Then the wave function [see
Fig.~\ref{fig:2} (b)] concentrates in the vicinity of
$r_{k_{\textrm{min}}}$ rather than at other $r_k.$ Its amplitude
almost four times as large as the value of the wave functions in
the next extremums. In the potential-free version the phase,
expectedly, reaches maximum at higher energy ($E=7.41$ MeV).
Again the wave function concentrates near $r_{k_{\textrm{min}}},$
although not as much as in the previous case.

\subsection{System $^3$H+$^3$H}

Now let us consider a collision of the two nuclei $^3$H with
opposite spins. In this case the norm kernel contains a complete
basis of the Pauli-allowed basis functions and their eigenvalues
which correspond to the two different values of the total spin:
$S=0$ and $S=1$,
\begin{eqnarray}
I_{^3\textrm{H}+^3\textrm{H}}({\bf R},{\bf
S})&=&I_{^3\textrm{H}+^3\textrm{H}}^{S=0}({\bf R},{\bf
S})D^0_{00}(\sigma)\nonumber
\\& &+I_{^3\textrm{H}+^3\textrm{H}}^{S=1}({\bf R},{\bf S})D^1_{00}(\sigma)
\label{kern_h3}.
\end{eqnarray}
Here $D^S_{MM'}(\sigma)$ are the Wigner $D$-functions, $\sigma$
is a set of spin variables.

As the total isospin of the system is equal to unity, the even
number of quanta, obviously, corresponds to the singlet states,
\begin{eqnarray*}
I_{^3\textrm{H}+^3\textrm{H}}^{S=0}({\bf R},{\bf S})&=&\cosh({\bf
R}\cdot{\bf S})-\cosh\left[{\frac{({\bf R}\cdot{\bf
S})}{3}}\right]\nonumber
\\& =&\sum_{k=1}^\infty
\left[1-\left(\frac{1}{3}\right)^{2k}\right] \frac{1}{(2k)!}({\bf
R}{\bf S})^{2k};
\end{eqnarray*}
while the odd number of quanta, to the triplet ones,
\begin{eqnarray*}
I_{^3\textrm{H}+^3\textrm{H}}^{S=1}({\bf R},{\bf S})&=&\sinh({\bf
R}\cdot{\bf
S})-3\sinh\left[\frac{({\bf R}\cdot{\bf S})}{3}\right]\nonumber \\
&=&\sum_{k=1}^\infty
\left[1-\left(\frac{1}{3}\right)^{2k}\right]\nonumber \\&&
\times\frac{1}{(2k+1)!}({\bf R}\cdot{\bf S})^{2k+1}.
\end{eqnarray*}
At that, the eigenvalues are given by
\begin{eqnarray*}
\Lambda_{\nu=2k}^{S=0}=1-\left(\frac{1}{3}\right)^{\nu},\qquad
\Lambda_{\nu=2k+1}^{S=1}=1-\left(\frac{1}{3}\right)^{(\nu-1)}.
\end{eqnarray*}
In contrast to the system n+$\alpha$, the eigenvalues with the
even and odd number of quanta are identical and tend to unity from
below. As to the minimal number of quanta allowed by the Pauli
principle, it increases by one comparing to the previous case.

\begin{figure}[e]
\includegraphics[width=8.6cm]{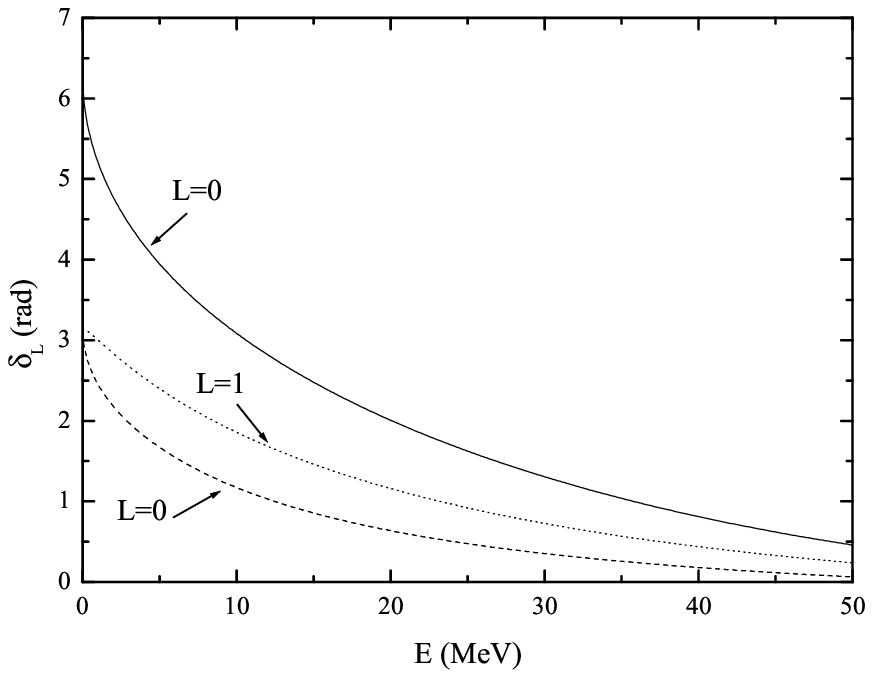}
\caption{\label{fig:4} Phases of the $^3$H+$^3$H scattering.
Solid curve: the phase obtained in the approximation of
zero-radius for nuclear force. Dotted and dashed curves: the
phases obtained by granting the Pauli principle only.}
\end{figure}

Again, as for the system n+$\alpha$ in the states with
$\Lambda_{2k}\\<1$, the scattering phase behavior determined only
by the influence of the antisymmetrized kinetic energy operator
corresponds to the repulsion of the clusters at small distances
between them. Because of the existence of the sole forbidden state
in both the channels, the scattering phases at zero energy are
counted off from $\pi$ (see Fig.~\ref{fig:4}). The singlet
scattering phase falls faster than the triplet one, because the
former has lower value of the angular momentum.
For the singlet state and $L=0$ a version with zero-range attractive potential
has also been considered, assuming that the interaction can be reproduced
by just one diagonal matrix element
\begin{eqnarray*}
<(2,0)|\hat{U}|(2,0)>=-34.76~\text{MeV}.
\end{eqnarray*}
Such model potential gives a correct value for the observed
threshold energy of the disintegration of the $^6$He nuclear
system into the channel $^3$H+$^3$H (12.3 MeV \cite{Ajzenberg})
and close to the experimental value the root-mean-square radius of
$^6$He nucleus, equal to 2.24 fm. The singlet phase of the elastic
scattering for the potential which assures an existence of one
bound state is counted off from $2\pi$ (Fig.~\ref{fig:4}). This
scattering phase is larger than the scattering phases obtained in
the potential-free version in the energy range being considered.

\subsection{System $\alpha$+$\alpha$}

Finally, let us address a well-known example of the scattering of
two $\alpha$-particles; or, in other words, let us consider $^8$Be
nuclear system in the $\alpha$-cluster model:
\begin{equation}
\label{alpha} I_{\alpha+\alpha}({\bf R},{\bf S})=
8\sinh^4\left[\frac{\left({\bf R}\cdot{\bf S}\right)}{4}\right].
\end{equation}
As seen from (\ref{alpha}), the norm kernel $I_{\alpha+\alpha}$
contains only the basis functions with the even number of quanta,
because the wave function of two identical bosons (which are the
$\alpha$-particles) should be symmetric with respect to
interchange of the clusters as a whole,
\begin{equation}
\label{kern_alpha} I_{\alpha+\alpha}({\bf R},{\bf S})=
\sum_{k=2}^\infty {1\over(2k)!}\left(1-\frac{4}{2^{2k}}\right)
({\bf R}\cdot{\bf S})^{2k}.
\end{equation}
An expression for the eigenvalues follows from (\ref{kern_alpha}),
\begin{eqnarray*}
\Lambda_{2k}=1-4\left(\frac{1}{4}\right)^{k},\qquad\Lambda_{2k+1}=0.
\end{eqnarray*}
$\Lambda_{2k}<1$ for any given number of quanta, and the minimal
allowed number of quanta takes maximum value
$\nu_{\textrm{min}}=4$ among all the cases considered in this
section. As a result, the Pauli principle leads to the repulsion
of clusters; and the behavior of the scattering phase is similar
to the above discussed one for the states with $L=0$.

\subsection{Comparison of the eigenvalues for different nuclear systems}

As has already been discussed in the previous section, at large
number of quanta the behavior of an effective potential related
to the antisymmetrization is determined by the expression
(\ref{range_eff}). In a general case, at small inter-cluster
distances an effective potential has rather a cumbersome form and
depends on several exponentially decreasing terms,
\begin{eqnarray}
\label{sum_eff} \Lambda_{\nu}^{(\lambda,\mu)}-1&=&
\sum_{j}\beta_j^{(\lambda,\mu)}(\nu)
\exp\left\{-\nu\ln\alpha_j^{(\lambda,\mu)}\right\},\\\nonumber&&
\alpha_j^{(\lambda,\mu)}>1.
\end{eqnarray}
However, in the problem of scattering of two $s$-clusters
at a  given $\nu$ the SU(3) irrep $(\nu,0)$ appears to be the only
possible representation; and in the sum (\ref{sum_eff}) only one
term remains. In Table~\ref{tab1} the parameters of an effective
interaction induced by the Pauli principle are presented for the
binary nuclear systems composed of the $s$-clusters.

\begin{table}
\caption{\label{tab1}Parameters of an effective interaction
related to the antisymmetrization,
$\Lambda_{k}-1=\beta\exp\left\{-k\ln\alpha\right\}$, $s$-cluster
systems.\\}
\begin{ruledtabular}
\begin{tabular}{ccccccc}
 & & &\multicolumn{2}{c}{$\nu=2k$} & \multicolumn{2}{c}{$\nu=2k+1$}\\
&$\nu_{\textrm{min}}$ &$\Lambda_{\nu_{\textrm{min}}}-1$& $\alpha$
& $\beta$& $\alpha$ &  $\beta$\\ \hline
n+$\alpha$ &1 &1/4 &16  &-1 & 16 &1/4\\
$^3$H+$^3$H&  2& -1/9& 9 & -1&  9&-1\\
$\alpha$+$\alpha$&  4& -1/4& 4& -4&  0&0
\end{tabular}
\end{ruledtabular}
\end{table}
By analyzing the data listed in Table~\ref{tab1}, some general
conclusions about dependence of the effective interaction caused
by the antisymmetrization effects on the number of nucleons in the
interacting clusters can be drawn.

As has already been mentioned in the previous section, one of the
quantities defining the range of the antisymmetrization operator
is the minimal number of quanta which corresponds to the first
non-vanishing eigenvalue. It is easy to understand that
$\nu_{\textrm{min}}$ increases with the number of nucleons of the
system under study, because the number of the occupied states
grows. Actually, in the system n+$\alpha$ the first allowed state
appears already at $\nu_{\textrm{min}}=1$, while in the case of
the two $\alpha$-particles it appears only at
$\nu_{\textrm{min}}=4$.

With the number of nucleons increasing, the parameter $\alpha$
defining the order of decrease for an effective potential also
essentially decreases. At the same time, the parameter $\beta$
rather increases that is an evidence of increasing the range of
antisymmetrization forces. A negative sign of $\beta$ indicates
that an effective interaction is repulsive one, except for the
states with the odd number of quanta in the system $^4$He+n where
it is attractive. Note, that in the latter case exactly those
basis states that correspond to the eigenvalues $\Lambda_{\nu}>1$
dominate in the wave function of $^4$He+n.

Finally, one more important parameter determining the intensity
of an effective potential related to the antisymmetrization is
the value of this potential at the minimal number of quanta. As
seen from Table~\ref{tab1}, the intensity of an effective
interaction is maximum in the states with
$\nu=\nu_{\textrm{min}}$  and increases with the number of
nucleons for the systems considered. Thus, a repulsive
interaction in the states with the even number of quanta for the
system $^4$He+n intensifies for the $^3$H+$^3$H nuclear system
and becomes even stronger in the system $\alpha$+$\alpha$.

\section{\label{sec:7}Multi-channel binary cluster systems}
\subsection{System $^6$He+$^6$He}

Cluster configuration $^6$He+$^6$He is a relatively simple
multi-channel system, by giving an example of which it is
possible to understand a role of the Pauli principle in
the formation of coupling between different channels.

Each of the clusters of the system $^6$He+$^6$He has open
$p$-shell; and, therefore, studying a collision of these clusters
it is natural to take account of their excitations accompanied by
the transition from the ground $0^+$ state to the 2$^+$ excited
state. This is a distinctive feature of the $^6$He+$^6$He nuclear
system in comparison with the above discussed. It consists, in
particular, in the fact that now at a given even value of the total
number of quanta $\nu=2k>8$ a basis of the Pauli-allowed states
with the total orbital angular momentum $L=0$ belongs to the five
SU(3) irreps with even symmetry indices $(\lambda,\mu)$:
$(2k-2,0),\,(2k,2)_1,\,(2k-4,4),\,(2k,2)_2,\,(2k+4,0).$ Notice
that the multiplets $(2k,2)_1$ and $(2k,2)_2$ have the same SU(3)
symmetry indices, but different eigenvalues $\Lambda_{(2k,2)_1}$
and $\Lambda_{(2k,2)_2}.$

As a result, the norm kernel $I_{^6\textrm{He}+^6\textrm{He}}$
for the states with $L=0$ takes the form
\begin{eqnarray*}
I_{^6\textrm{He}+^6\textrm{He}}&=&\sum_{k=2}^\infty
\Lambda_{(2k-2,0)}\psi_{(2k-2,0)}\tilde{\psi}_{(2k-2,0)}\\
& & +\sum_{k=3}^\infty
\Lambda_{(2k,2)_1}\psi_{(2k,2)_1}\tilde{\psi}_{(2k,2)_1}\\ & &+
\sum_{k=3}^\infty
\Lambda_{(2k-4,4)}\psi_{(2k-4,4)}\tilde{\psi}_{(2k-4,4)}\\ & &+
\sum_{k=4}^\infty
\Lambda_{(2k,2)_2}\psi_{(2k,2)_2}\tilde{\psi}_{(2k,2)_2}\\ & &+
\sum_{k=5}^\infty
\Lambda_{(2k+4,0)}\psi_{(2k+4,0)}\tilde{\psi}_{(2k+4,0)}.
\end{eqnarray*}
These eigenfunctions, along with their eigenvalues, were defined
in \cite{Fewbody2}. Hence, here we will restrict ourselves with
only those results from \cite{Fewbody2} which directly relate the
problem discussed.

\begin{figure}[e]
\includegraphics[width=8.6cm]{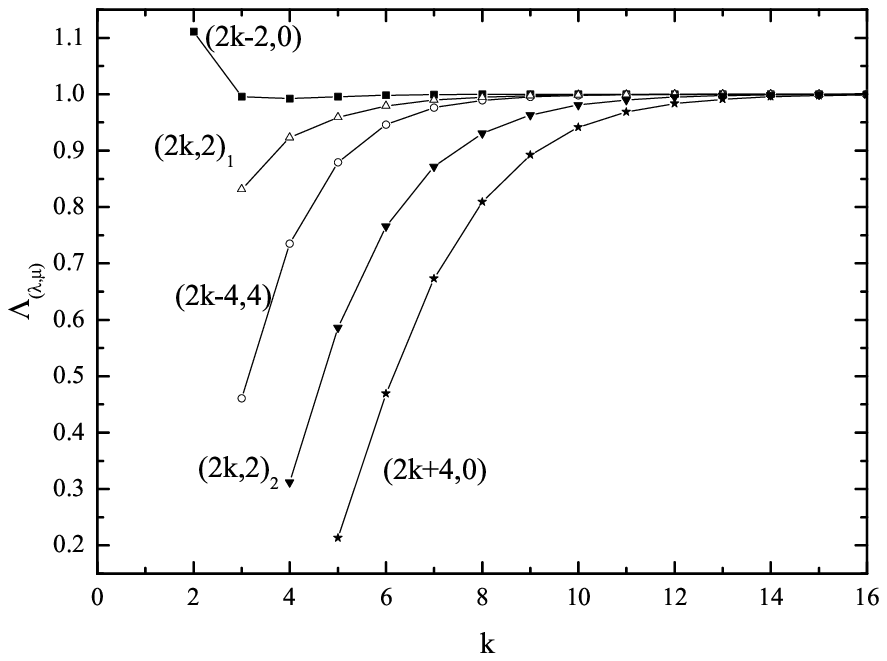}
\caption{\label{fig:7} Eigenvalues $\Lambda_{(\lambda,\mu)}$ of
the norm kernel for the system $^6$He+$^6$He versus the number of
quanta $k$. SU(3) symmetry indices $(\lambda,\mu)$ are shown near
the curves.}
\end{figure}

Such a plenty of the irreducible representations puts a question
of their classification with the aim of determining the most
important irreps, at least, for the low above-threshold energies.
First, let us present  in Fig.~\ref{fig:7} the dependence of the
eigenvalues belonging to the five different branches (according to
their definition introduced in Section~\ref{sec:2}) on $k$.
As seen from this figure, all the eigenvalues except
$\Lambda_{(2,0)}$ are less than unity; and, therefore, generate an
effective repulsive potential. The largest eigenvalues belong to
the branch $(2k-2,0);$ while the smallest ones, to the branch
$(2k+4,0).$ In the states of the latter branch a cluster
repulsion caused by the action of the Pauli principle is maximal,
as well as the range of the antisymmetrization effects. Only if
$k\geq14,$ eigenvalues are close to unity. Besides, this branch
starts with $k=5,$ i.e. later than the others. The repulsion in
the states of the branch $(2k,2)_2$, for which the minimal number
of $k$ equals 4, is somewhat less intensive; and its eigenvalues
can be set to unity, if $k\geq13.$ The repulsion for the branches
$(2k-4,4)$ and $(2k,2)_1$, which appear at $k=3$, is even less
pronounced. The eigenvalues of these branches are rather close to
unity, if $k\geq10.$ The fact, that in the absence of degeneracy
the higher U(3)-symmetry (the larger the eigenvalues of the
second-order Casimir operator of U(3)-group), the smaller the
eigenvalues, attracts attention.

Of course, the most remarkable feature of the basis of the
Pauli-allowed states for the $^6$He+$^6$He system is that this
basis corresponds to the five different channels. Above some
threshold energy ($E=3.6$ MeV) all these channels are open. But
there is such energy range in the continuous part of the spectrum
where two or only one channel is open. The influence of the Pauli
principle on the system considered manifests itself in making all
the five channels coupled at small inter-cluster distances. The
radius of this domain is determined by the requirement that on
its border the five different eigenvalues of the allowed states
are almost equal to unity. Below we shall specify at what real
values of $k$ (and, therefore, $r_k$) it occurs. As soon as all
the eigenvalues approach unity, a unitary transformation from the
SU(3) basis to the $l$-basis, which allows us to make the coupling
of different channels of the latter basis via the kinetic energy
operator vanish, becomes possible \cite{Fewbody2}.

Provided that the eigenvalues of different SU(3) branches are not
identical, coupling of the channels via matrix elements of the
kinetic energy operator directly results in an appearance of the
off-diagonal elements of the scattering $S$-matrix; and, hence, in
an occurrence of the inelastic processes in the collision of the
two $^6$He nuclei. Certainly, a potential energy of cluster
interaction also can influence the inelastic scattering
cross-sections. However, as before, we shall restrict our
analysis to only that contribution to the nuclear reaction
cross-section which is made by the effects of exchange of the
nucleons belonging to the different clusters. A simple potential
energy operator used here does not couple different channels.

\begin{figure}[e]
\includegraphics[width=8.6cm]{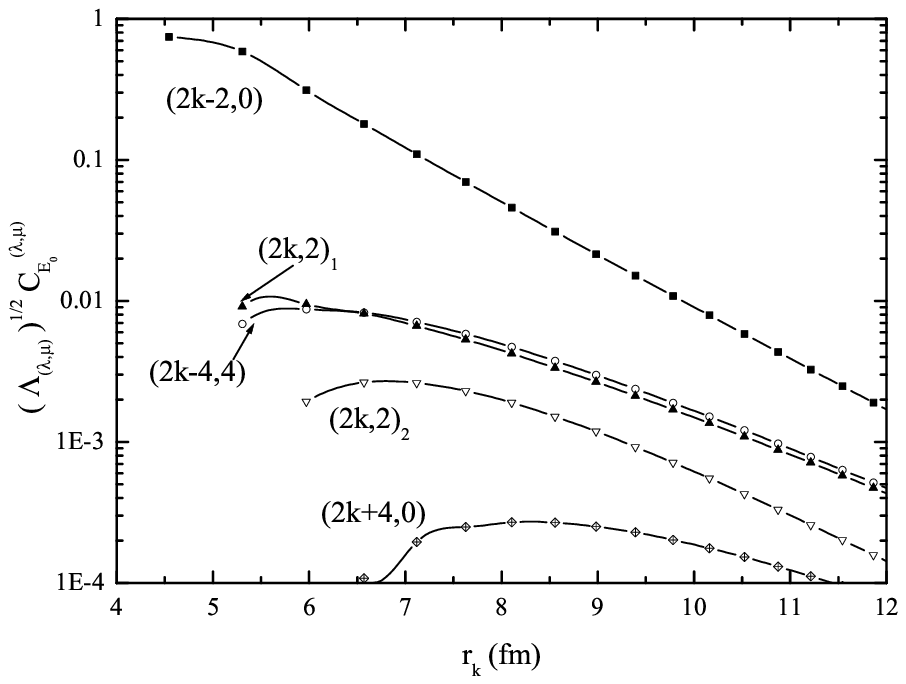}
\caption{\label{fig:8} Ground state of $^{12}$Be=$^6$He+$^6$He:
coefficients
$\sqrt{\Lambda_{(\lambda,\mu)}}$$C^{(\lambda,\mu)}_{E_0}(r_k)$ of
the w.f. expansion in the SU(3) basis, half-logarithmic scale.
SU(3) symmetry indices $(\lambda,\mu)$ are shown near the curves.}
\end{figure}

An information about the intensity of repulsion in the states
belonging to the different branches is provided by the expansion
coefficients (Fig.~\ref{fig:8}) of the ground state wave function
of $^{12}$Be=$^6$He+$^6$He calculated in the zero-range
approximation for nuclear force \cite{Fewbody2}.
We assume that the interaction can be reproduced by just two
diagonal matrix elements in the SU(3) representation $(2k-2,0)$,
i.e.
\begin{widetext}
\begin{eqnarray*}
\langle(2k-2,0)|U|(2k'-2,0)\rangle=\left\{
\begin{array}{cc} U_0=- 44.2 \text{ MeV} & \text{ if } k=k'=2 \\
U_1=- 28.7 \text{ MeV} & \text{ if } k=k'=3 \\0  & \text{
otherwise }   \end{array} \right.
\end{eqnarray*}
\end{widetext}
These values were fitted to the experimental values of the r.m.s.
radius of $^{12}$Be ($2.59\pm0.06$ fm \cite{Tanihata}) in its
ground state, and the $^6$He$+^6$He decay threshold energy
($10.11$ MeV \cite{Freer2}). The oscillator length was fixed to
$1.37$ fm.

Comparing the coefficients of different branches, we come to a
conclusion that the same inequality holds both for them and for
their eigenvalues,
$$\Lambda_{(2k-2,0)}\geq\Lambda_{(2k,2,)_1}\geq\Lambda_{(2k-4,4,)}
\geq\Lambda_{(2k,2,)_2}\geq\Lambda_{(2k+4,0)},$$ with the only
small exception: the coefficients of the branch $(2k-4,4,)$ are
somewhat larger than the coefficients of the branch $(2k,2)_1,$
if $r_k>6.5$ fm.
\begin{figure}[e]
\includegraphics[width=8.6cm]{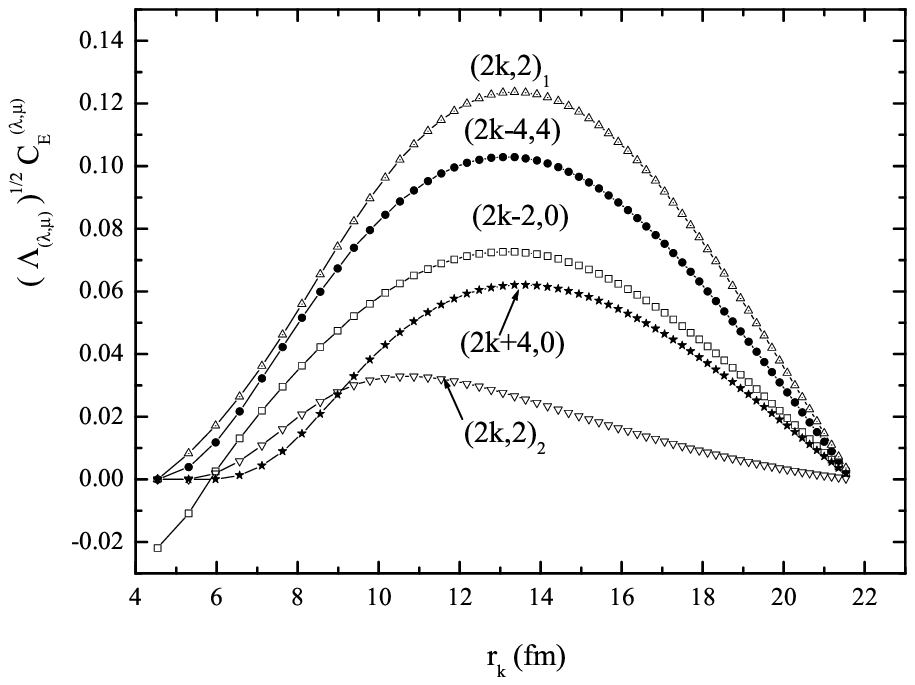}
\caption{\label{fig:9} Coefficients
$\sqrt{\Lambda_{(\lambda,\mu)}}C^{(\lambda,\mu)}_E(r_k)$ of the
expansion of the continuum states of $^{12}$Be=$^6$He+$^6$He in
the SU(3) basis at $E=0.885$ MeV. SU(3) symmetry indices
$(\lambda,\mu)$ are shown near the curves.}
\end{figure}

Let us consider the wave function for one more state
(Fig.~\ref{fig:9}) obtained by diagonalization of the Hamiltonian
on the SU(3) basis states, with the maximal value of $k$ being
equal to 64. This state was chosen in such a way to make its
energy $E=0.885$ MeV be above the threshold of the $^{12}$Be
decay into two $^6$He nuclei in their ground state ($E=0$); but
less than the threshold energy ($E=1.8$ MeV) of the decay of
$^{12}$Be into the channel, one of the clusters being in $2^+$
excited state \footnote{Its experimental width is only about 113
keV, so we treat it as a bound state here.}. As long as $r_k<6$
fm, the behavior of the coefficients is determined by the
intensity of repulsion in the corresponding SU(3) branches, like
in the g.s. wave function. However, then a rearrangement of their
values occurs. The coefficients of the irreps $(2k,2)_1$ appear
to be leading ones, and they are followed by the irreps $(2k-4,4)$
and $(2k-2,0).$ Enumerated coefficients correlate in such a way
that by projecting on the states of $l$-basis we arrive at the
basis functions of the open channel. At large $r_k$ the basis
functions of the irreps $(2k,2)_2$ contain only those functions
of the $l$-basis which correspond to the closed channels. That is
why the expansion coefficients belonging to the branch $(2k,2)_2$
exponentially tend to zero with $r_k$ increasing.
\begin{figure}[e]
\includegraphics[width=8.6cm]{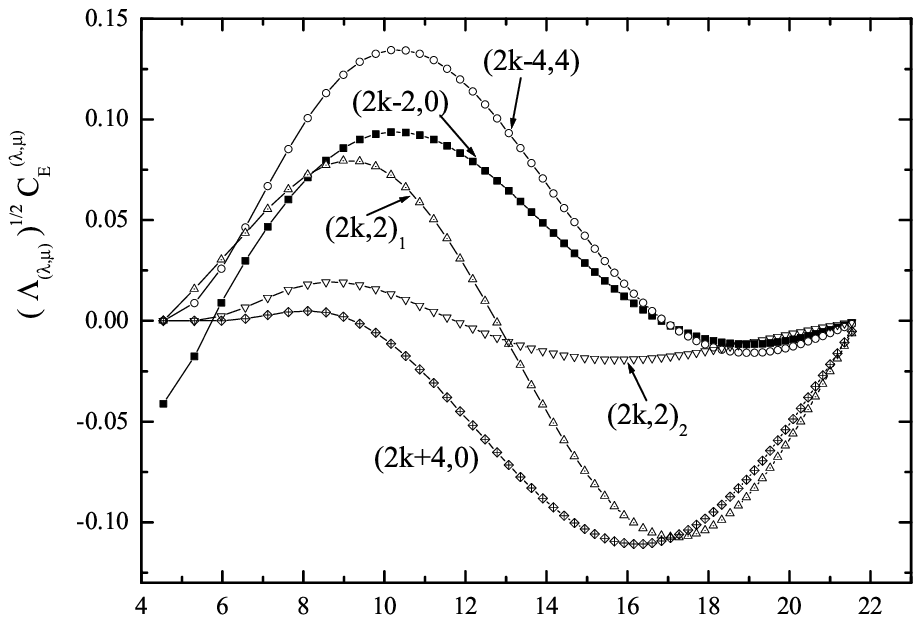}
\caption{\label{fig:10} Coefficients
$\sqrt{\Lambda_{(\lambda,\mu)}}C^{(\lambda,\mu)}_E(r_k)$ of the
expansion of the continuum states of $^{12}$Be=$^6$He+$^6$He in
the SU(3) basis at $E=3.3$ MeV. SU(3) symmetry indices
$(\lambda,\mu)$ are shown near the curves.}
\end{figure}

Finally, turn to the state with the energy $E=3.3$ MeV
(Fig.~\ref{fig:10}). This state is above the threshold for the
decay of $^{12}$Be into the two $^6$He nuclei; provided that one
of them being in the ground state, while the other being excited.
In this energy region all the states with zero angular momentum
are two-fold degenerate, as the two channels are open. For the
expansion coefficients the same behavior is observed as in the
previous case, granting the
 fact that in this case the coefficients are nonzero  for only those
SU(3) irreps which contain the basis functions of the two open
channels.

Multi-channel situation make analysis of the scattering
eigenphases difficult. The question is from what value the
eigenphases should be counted off at zero threshold energy of the
corresponding channel, to make the eigenphases vanish at high
energy $E$. At first, notice that after a transformation to the
$l$-basis the total number of oscillator quanta $2k$ used for the
SU(3) symmetry indices of different branches is appropriate to
represent as a sum $2k=2k'+l,$ where $l$ is even angular momentum
of cluster relative motion, $k'$ is the number of radial quanta.
Let us begin with the fact that the first (regarding the number
of quanta) Pauli-allowed state possesses SU(3) symmetry $(2,0).$
After a transformation to the $l$-basis it appears to contain the
functions
\begin{eqnarray*}
\Phi_2^{(0,0,0)},~~\Phi_2^{(2,2,0)},~~
\frac{1}{\sqrt{2}}\left\{\Phi_2^{(2,2,0)}+\Phi_2^{(0,2,2)}\right\},
\text{ and } \Phi_2^{(2,2,2)}.
\end{eqnarray*}
 If the angular momentum $l$
of the relative motion equals zero, the Pauli-forbidden states are
$$\Phi_0^{(0,0,0)},~~\Phi_1^{(0,0,0)}; \text{ and }
\Phi_0^{(2,2,0)},~~\Phi_1^{(2,2,0)}.$$ Hence, the eigenphases of
these channels are appropriate to be counted off from $2\pi$ at
zero energy in the corresponding channel. Even though $l=2,$ then
in the channel where only one of the two $^6$He nuclei is excited
the two states are forbidden \footnote{At $k=0$ the states with
$l=2$ do not exist.}:
$$\frac{1}{\sqrt{2}}\left\{\Phi_1^{(2,0,2)}+\Phi_1^{(0,2,2)}\right\}
\text{ and }
\frac{1}{\sqrt{2}}\left\{\Phi_1^{(2,0,2)}-\Phi_1^{(0,2,2)}\right\}.$$
Therefore, in this channel the eigenphase will also be counted
off from $2\pi$. One more forbidden state, $\Phi_1^{(2,2,2)}$,
specifies counting off from $\pi$ for the eigenphase in the
channel with both $^6$He being excited. At last, the channel
$l=4$ also has only one forbidden state $\Phi_2^{(2,2,4)},$ as
this function appears only at $k=3$ along with the functions of
the irreps $(6,2)_1$ and $(2,4).$ Hence, the eigenphase
$\delta_{l=4}$ will be counted off from $\pi.$ To summarize, note
that there are eight different forbidden states of the $l$-basis.
Besides, because of existence of the sole bound state in the
zero-approximation for nuclear force, all the eigenphases should
be moved up by $\pi$ more.
\begin{figure}[e]
\includegraphics[width=8.6cm]{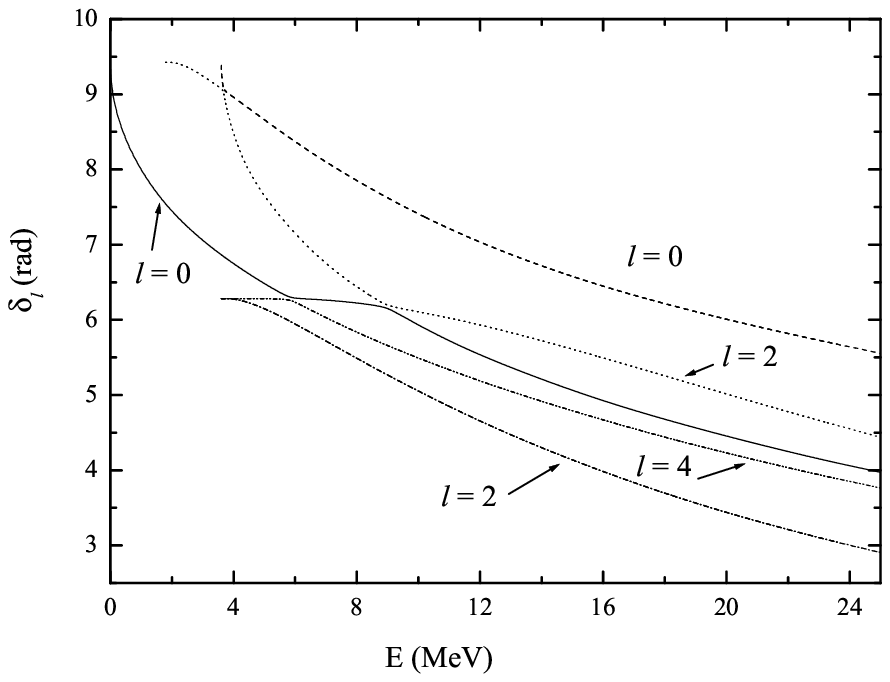}
\caption{\label{fig:11} Eigenphases $\delta_l(E)$ of the
$^6$He+$^6$He system obtained in the approximation of zero-range
for nuclear force. Values of the angular momentum $l$ of the
cluster relative motion are shown near the curves.}
\end{figure}

Important information about the multi-channel continuous spectrum
of the $^6$He+$^6$He system is provided by its five (according to
the maximal number of the open channels) eigenphases, presented
in Fig.~\ref{fig:11}. Just above the corresponding threshold
$E_{\textrm{thr}},$ where one or another channel opens, the
eigenphases $\delta_l$ obey the law
$$\delta_l(E)=\delta_l(E_{\textrm{thr}})+\text{const}E^{l+1/2}.$$
Here $\delta_l(E_{\textrm{thr}})$ is the value divisible by
$\pi.$  In Fig.~\ref{fig:11} the three quasi-intersections of the
phase curves are seen. Eigenphases can not intersect, because the
fact of intersection would contradict the unicity theorem for a
solution of a wave equation. After the intersection point each of
the eigenphases moves in the direction along which the other
eigenphase moved before the intersection point. Fall of the
eigenphases with the energy indicates a repulsion due to the
antisymmetrization effects, and is not compatible with the
assumption that any resonance exists in the system in question.

\subsection{$^4$He+$^8$He and $^6$He+$^6$He clustering: coupled-channel approach}

Along with the clustering of $^{12}$Be into the channels with both
$^6$He nuclei being in the ground or excited state, let us
consider also the $^4$He+$^8$He cluster structure. The latter
allows an excitation of the $^8$He nucleus to its $2^+$ state.
Taking into account the  $^4$He+$^8$He clustering results in the
important corrections to the results of calculations with due
regard of the $^6$He+$^6$He clustering only.

First of all, the number of the allowed states increases and an
additional SU(3) degeneration appears. As a result, the number of
the channels with the total angular momentum $L=0$, total spin
$S=0$, and isospin $T=2$ grows to seven. The $^4$He+$^8$He
clustering provides for the two additional branches of the basis
states having SU(3) symmetry $(2k-2,0)$ and $(2k,2).$ Basis
states of different cluster configurations with the same SU(3)
symmetry indices are not orthogonal, while the two states $(2,0)$
are identical. Hence, we considering the two clustering modes
simultaneously; the new basis states (the eigenfunctions of the
norm kernel for the two coupled cluster configurations)
representing a superposition of the basis functions of the
$^4$He+$^8$He and $^6$He+$^6$He channels are obtained.

\begin{figure*}[e]
\includegraphics[width=17.6cm]{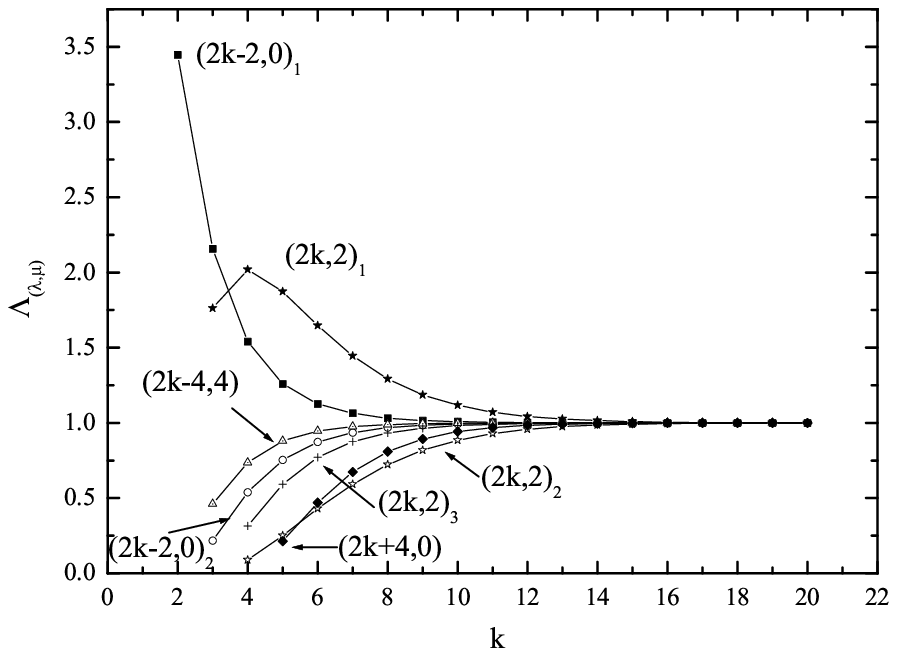}
\caption{\label{fig:12} Eigenvalues $\Lambda_{(\lambda,\mu)}$ of
the norm kernel for the $^{12}$Be system
versus the number of quanta $k$. SU(3) symmetry indices
$(\lambda,\mu)$ are shown near the curves.}
\end{figure*}

Let us start with the Hilbert-Schmidt expansion of the new norm
kernel $I_{(8,4)+(6,6)}$ for the states with $L=0.$ Seven
different sums are written down in a descending order of the
eigenvalues $\Lambda_{(\lambda,\mu)},$
\begin{eqnarray*}
I_{(8,4)+(6,6)}=&&\sum_{k=2}^\infty \Lambda_{(2k-2,0)_1}
\psi_{(2k-2,0)_1}\tilde{\psi}_{(2k-2,0)_1}\\&&+\sum_{k=3}^\infty
\Lambda_{(2k,2)_1}\psi_{(2k,2)_1}\tilde{\psi}_{(2k,2)_1}\\&&
+\sum_{k=3}^\infty
\Lambda_{(2k-4,4)}\psi_{(2k-4,4)}\tilde{\psi}_{(2k-4,4)}\\&&+
\sum_{k=3}^\infty \Lambda_{(2k-2,0)_2}
\psi_{(2k-2,0)_2}\tilde{\psi}_{(2k-2,0)_2}\\&&+\sum_{k=4}^\infty
\Lambda_{(2k,2)_2}\psi_{(2k,2)_2}\tilde{\psi}_{(2k,2)_2}\\&&
+\sum_{k=4}^\infty
\Lambda_{(2k,2)_3}\psi_{(2k,2)_3}\tilde{\psi}_{(2k,2)_3}\\&&+
\sum_{k=5}^\infty
\Lambda_{(2k+4,0)}\psi_{(2k+4,0)}\tilde{\psi}_{(2k+4,0)}.
\end{eqnarray*}
Now the states $(2k-2,0)$ become two-fold degenerate, while the
degree of degeneracy for the states $(2k,2)$ raises to three. The
dependence on the number of quanta of the eigenvalues belonging to
the different branches  is shown in Fig.~\ref{fig:12}.
Its remarkable feature is that now the eigenvalues of the two
branches, $\Lambda_{(2k-2,0)_1}$ and $\Lambda_{(2k,2)_1}$, exceed
unity. An effective potential related to antisymmetrization in
the states of the aforementioned branches corresponds to the
attraction. This directly affects the structure of the g.s. wave
function of the $^{12}$Be nucleus, because now the expansion
coefficients belonging to the SU(3) irreps $(2k-2,0)_1$ dominate
in the expression for the wave function of the $^{12}$Be ground
state. The basis functions of these irreps on equal footing
contain both the states of the $^6$He+$^6$He and $^4$He+$^8$He
clustering.

Now in order to reproduce the threshold energy (8.95 MeV
\cite{freer}) of the $^{12}$Be break-up into $^4$He and $^8$He in
their ground states  and the root-mean-square radius of the
$^{12}$Be nucleus, $r_{\textrm{rms}}=2.66$ fm, the parameter $U_0$
of the model zero-range potential should be reduced in absolute
magnitude to the value of 42.2 MeV; while the parameter $U_1$
should even be set to zero. Such a change of the potential
parameters seems to be natural, because the basis of the
$^6$He+$^6$He configuration is supplemented with the basis states
of the $^4$He+$^8$He configuration, with a contribution of the
latter ones to the energy of the ground state being considerable.

\begin{figure}[e]
\includegraphics[width=8.6cm]{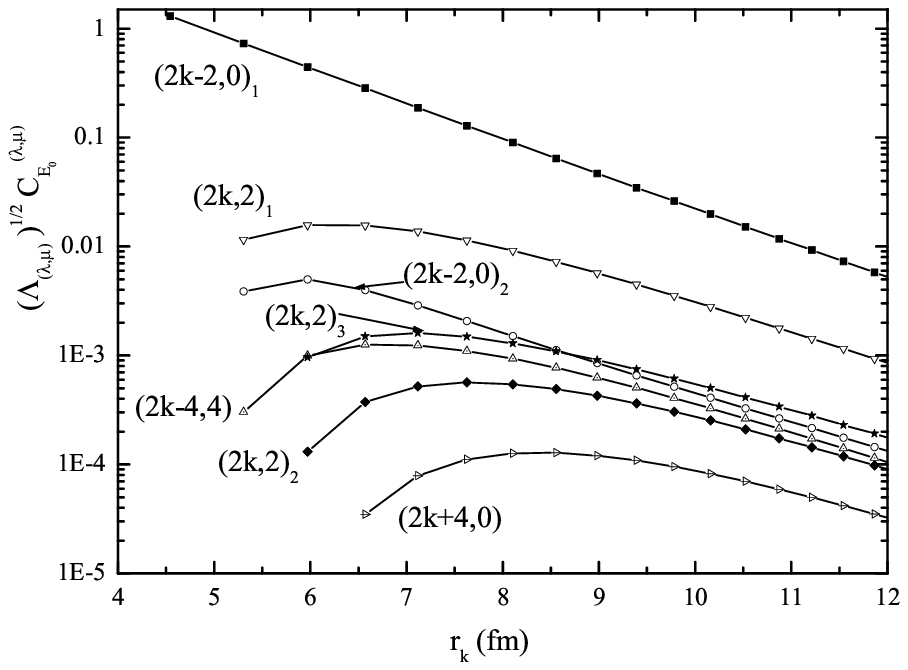}
\caption{\label{fig:13} Ground state of the $^{12}$Be system:
coefficients
$\sqrt{\Lambda_{(\lambda,\mu)}}C^{(\lambda,\mu)}_{E_0}(r_k)$ of
the w.f. expansion in the SU(3) basis, half-logarithmic scale.
SU(3) symmetry indices $(\lambda,\mu)$ are shown near the curves.}
\end{figure}

The coefficients $\sqrt{\Lambda_{(\lambda,\mu)}}
C_{E_{0}}^{(\lambda,\mu)}(r_k)$ are presented in
Fig.~\ref{fig:13}.
With the small exception, the smaller are the eigenvalues of the
basis functions, the smaller are the corresponding coefficients.

\begin{figure}[e]
\includegraphics[width=8.6cm]{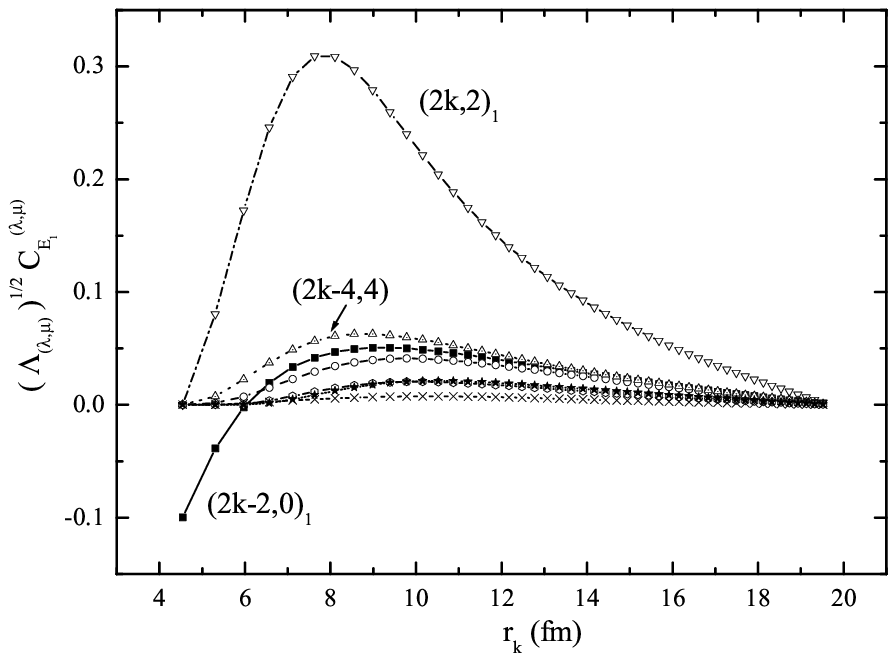}
\caption{\label{fig:14} Coefficients
$\sqrt{\Lambda_{(\lambda,\mu)}}C^{(\lambda,\mu)}_{E_1}(r_k)$ of
the expansion of the second bound state of $^{12}$Be system
in the SU(3) basis at $E_1=-0.386$ MeV. SU(3) symmetry indices
$(\lambda,\mu)$ are shown near the curves.}
\end{figure}

The chosen version of the potential provides for an existence of
the second bound state at the energy $E_1=-0.386$ MeV. The
coefficients
$\sqrt{\Lambda_{(\lambda,\mu)}}C_{E_1}^{(\lambda,\mu)}(r_k)$ are
presented in Fig.~\ref{fig:14}. The coefficients of the irreps
$(2k,2)_1$ appear to be dominating for this state, while the
coefficients of the  $(2k-2,0)_1$ irreps take the second place
where they compete with the coefficients of the irreps
$(2k-4,4).$ The contribution of the other coefficients again
correlates with the magnitude of their eigenvalues.

\begin{figure}[e]
\includegraphics[width=8.6cm]{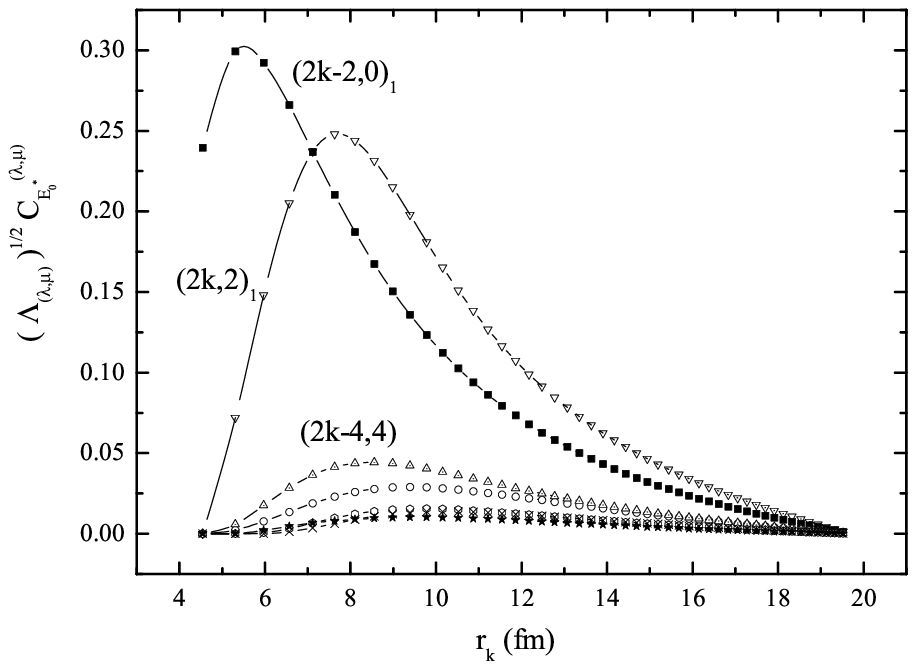}
\caption{\label{fig:15} Ground state of the $^{12}$Be system:
coefficients
$\sqrt{\Lambda_{(\lambda,\mu)}}C^{(\lambda,\mu)}_{E_0^*}(r_k)$ of
the w.f. expansion in the SU(3) basis. SU(3) symmetry indices
$(\lambda,\mu)$ are shown near the curves.}
\end{figure}

Moreover, now even without any cluster interaction potential
generated by the nucleon-nucleon forces the bound state of the
$^{12}$Be nuclear system, with the energy $E_0^*=-0.75$ MeV and
the root-mean-square radius $r_{\textrm{rms}}=4$ fm, appears. In
Fig.~\ref{fig:15} the expansion coefficients of the wave function
for this state are shown. Those of the coefficients that
correspond to the irreps $(2k-2,0)_1$ and $(2k,2)_1$ possess the
largest values among all the coefficients at small $r_k$. The
bound state appeared due to the attraction for which the two
latter branches of the SU(3) irreps are responsible.
For simplicity, the five threshold energies are assumed to be
equal. But even then the question of the correct value assignment
for each of the seven eigenphases at zero energy remains to be
clarified. An answer to this question is reduced to the
calculation of the total number of the forbidden states of the
$l$-basis . The $^8$He+$^4$He configuration supplies the three
forbidden states. One of them corresponds to the channel with
$^8$He and $^4$He clusters being in their ground states, zero
angular momentum of their relative motion and zero number of
quanta, i.e. $k=0.$ The other forbidden state corresponds to the
same channel, but with $k=1.$ The last forbidden state belongs to
the channel with $k=1$ and $^8$He nucleus excited.

\begin{figure*}[e]
\includegraphics[width=8.6cm]{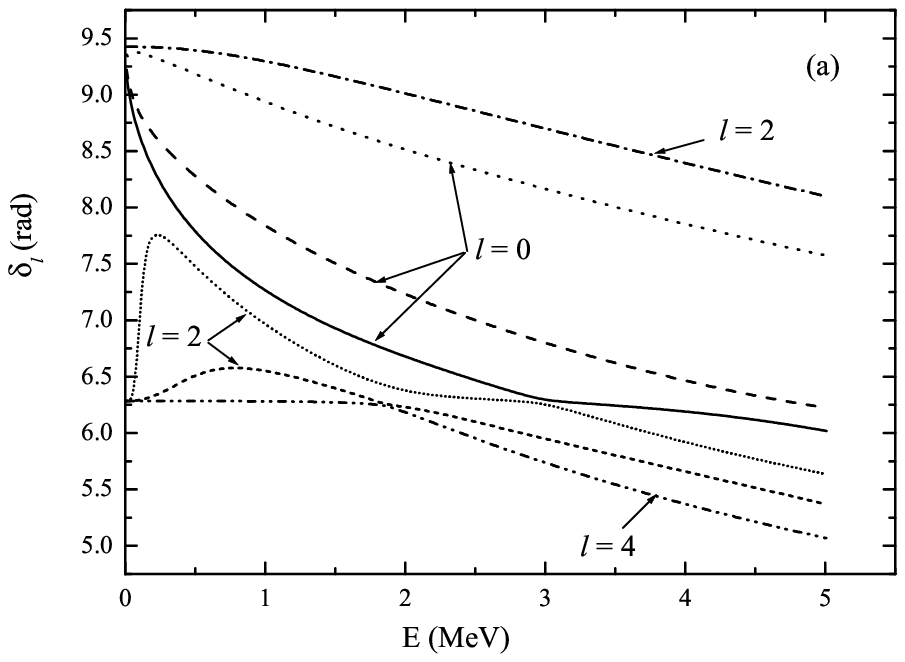}
\includegraphics[width=8.6cm]{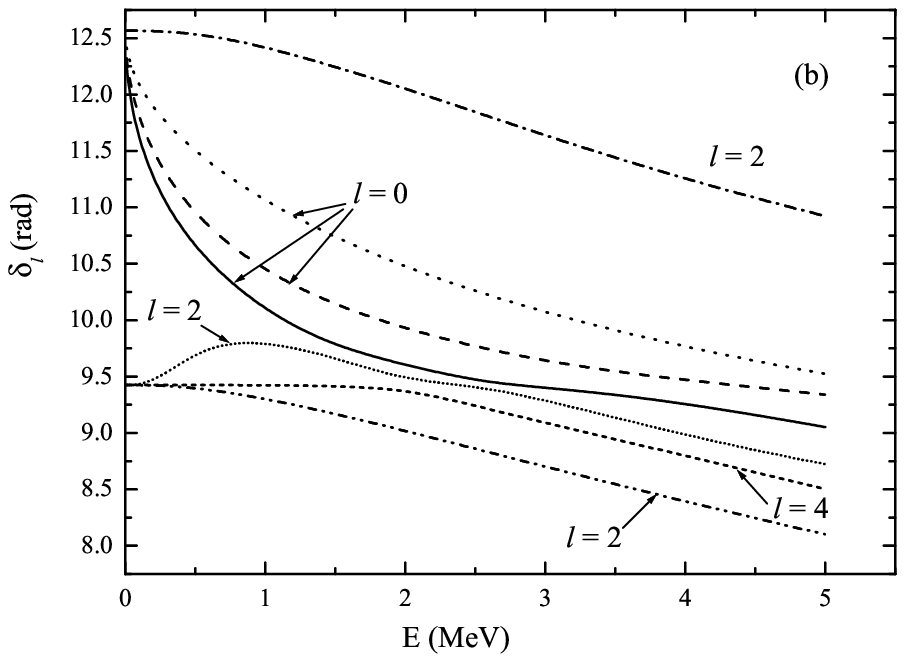}
\caption{\label{fig:16} Eigenphases $\delta_l(E)$ of the
$^{12}$Be system.
(a) The eigenphases obtained by granting the antisymmetrization
effects only. (b) The eigenphases obtained in the zero-range
approximation for nuclear force. Values of the angular momentum
$l$ of the cluster relative motion are shown near the curves.}
\end{figure*}

Granting all the above-mentioned concerning the forbidden states
of the $^6$He+$^6$He clustering, we arrive at the conclusion
about an existence of the eleven forbidden states. As regards the
magnitude of the eigenphases at zero energy, with due regard of
the antisymmetrization effects only the eigenphases of the three
channels with $l=0$ begin with $3\pi$ like the eigenphase for one
of the three channels with $l=2.$ The eigenphases for the other
channels with $l=2$ and the eigenphase of the channel with $l=4$
begin with $2\pi.$ The energy dependence for all the eigenphases
is presented in Fig.~\ref{fig:16}.

\begin{figure*}[e]
\includegraphics[width=17.8cm]{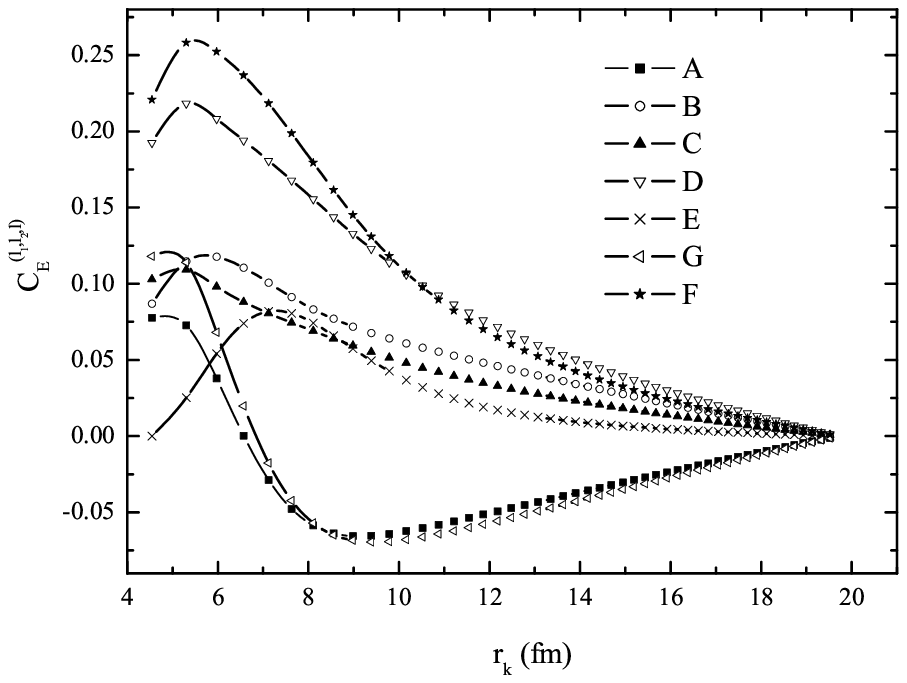}
\caption{\label{fig:17}Coefficients $C^{(l_1,l_2,l)}_E(r_k)$ of
the expansion of the continuum states of the $^{12}$Be system
in the
$l$-basis at $E=0.22$ MeV. $^6$He+$^6$He clustering:
line A -- $l_1=l_2=l=0$; line B -- $l_1=l_2=2$, $l=0$;
line C -- $l_1=2 (0),~l_2=0 (2),~l=0$; line D -- $l_1=l_2=l=2$;
line E -- $l_1=l_2=2,~l=4$;
$^8$He+$^4$He clustering: line G -- $l_1=l_2=l=0$;
 line F -- $l_1=l=2,~l_2=0$.}
\end{figure*}

Let us remind that for the $^6$He+$^6$He clustering all the
eigenphases decrease monotonically with the energy. For the case
of simultaneous consideration of the $^6$He+$^6$He and
$^8$He+$^4$He configurations, but without a nucleon-nucleon
potential [Fig.~\ref{fig:16} (a)], two of the seven eigenphases
with the angular momentum $l=2$ of the relative cluster motion
first ascend, reach maximum within the energy range up to 1 MeV,
and only after that begin to decrease monotonically. We have
already observed such a behavior for  $\alpha+$n scattering phase
with the angular momentum $L=1$. To understand, whether it is
possible to conclude about resonances owing to these maxima, let
us consider the wave function in the continuum for the energy of
0.22 MeV obtained by diagonalization of the Hamiltonian (see
Fig.~\ref{fig:17}). The chosen energy is close to that one at
which one of the eigenphases has a pronounced peak. In this wave
function the states corresponding to the aforementioned channels
with $l=2$ prevail that is compatible with the assumption about
an existence of the resonance.

Inclusion of the zero-range potential pulls down the resonance
under the break-up threshold of $^{12}$Be into $^8$He+$^4$He
[Fig.~\ref{fig:16} (b)]. As a result, along with the ground state
of the $^{12}$Be nucleus ($E_0=-8.95$ MeV), the excited state
appears at the energy $E_1=-0.386$ MeV. Further behavior of the
eigenphases with the energy increasing is similar to that which
has already been discussed for the $^6$He+$^6$He clustering.

\section{Conclusions}

The Pauli principle influence on the structure of the continuum
states of the compound-systems populated at the intermediate
stage of collisions between light nuclei was studied within the
algebraic version of resonating group method. The exchange
effects generated by the antisymmetrization operator were
analyzed on the ground of the discrete representation of the
complete basis of the Pauli-allowed many-particle
harmonic-oscillator states classified with the use of SU(3)
symmetry indices. The eigenvalue problem for the norm kernel (the
overlap integral of the antisymmetric generating functions of the
Hill-Wheeler method) was reduced to a solution of the degenerate
integral equations in the Fock-Bargmann space.

The influence of the Pauli exclusion principle on the collision
of clusters was shown to be reducible to three factors which
affect the dynamics of the cluster-cluster interaction. Firstly,
elimination of the forbidden states drastically increases the
maximum amplitude of the scattering phase that may be simulated
by a repulsive potential at small inter-cluster distances. The
larger is the number of the forbidden states, the larger should be
the intensity and the radius of such a model potential. Secondly,
out of the core (in the region where the eigenvalues of the
Pauli-allowed states differ from unity) some additional effective
potential (repulsive or attractive) appears. The latter can
significantly affect the scattering phase behaviour. Finally,
decreasing or increasing the centrifugal potential occurs in the
same region. It also influences the phase shift, especially, at
high energy. The inter-cluster distances, within which the Pauli
principle is important, depend on the cluster structure and can
be several times more than the radius of the cluster-cluster
interaction induced by the nucleon-nucleon potential.

If there are several open channels, exchange of the nucleons which
belong to the colliding clusters forms the cross-sections of the
inelastic scattering.
The eigenphases of the multi-channel systems that define nature of
the inelastic collisions have been calculated; and the fact of
quasi-intersection of the eigenphases was established.

A considerable intensification of the antisymmetrization effects
is observed in the case when different cluster configurations
(such as, for example, $^6$He+$^6$He and $^4$He+$^8$He which are
actual for the $^{12}$Be compound nucleus) are taken into account
simultaneously. This phenomenon relates to an appearance of the
new branches of excitation with the especially large, greater
than unity, eigenvalues of the allowed states. As a result, an
effective attraction induced by the antisymmetrization effects
appears to be strong enough to ensure an existence both the bound
state and the resonance even without a participation of the
nucleon-nucleon interaction between nucleons of different
clusters.

\begin{acknowledgments}
The authors would like to express their gratitude to
V.~S.~Vasilevsky for valuable discussions.
\end{acknowledgments}

\end{document}